\documentclass{article}

\usepackage{PRIMEarxiv}

\usepackage[utf8]{inputenc} 
\usepackage[T1]{fontenc}    
\usepackage{hyperref}       
\usepackage{url}            
\usepackage{booktabs}       
\usepackage{amsfonts}       
\usepackage{nicefrac}       
\usepackage{microtype}      
\usepackage{lipsum}
\usepackage{fancyhdr}       
\usepackage{graphicx}       
\graphicspath{{media/}}     
\usepackage{wrapfig}
\usepackage{amsmath}
\usepackage{amsfonts}
\usepackage{graphicx}
\usepackage{parskip}
\usepackage{comment}
\usepackage{csquotes}
\usepackage{geometry}
\usepackage{array}
\usepackage{booktabs}
\usepackage[most]{tcolorbox}
\usepackage{wrapfig}
\usepackage{tabularx}
\usepackage[font=small,labelfont=bf,justification=centering]{caption}

\newcolumntype{C}{>{\centering\arraybackslash}X}

\AtBeginDocument{
    \setlength{\abovedisplayskip}{1.25em}
    \setlength{\belowdisplayskip}{1.25em}
    \setlength{\abovedisplayshortskip}{0.75em}
    \setlength{\belowdisplayshortskip}{0.75em}
}

\usepackage[
backend=biber,
style=authoryear,
]{biblatex}

\newcommand\blfootnote[1]{%
  \begingroup
  \renewcommand\thefootnote{}\footnote{#1}%
  \addtocounter{footnote}{-1}%
  \endgroup
}

\def\sym#1{\ifmmode^{#1}\else\(^{#1}\)\fi}

\newcommand{\flop}{\, \textrm{FLOP}}

\newcommand{\flops}{\, \textrm{FLOP}/\textrm{s}}
\newcommand{\flopyr}{\, \textrm{FLOP}/\textrm{year}}

\newcommand{\oom}{\, \textrm{OOM}}

\title{Explosive growth from AI automation: A review of the arguments}
\date{July 2023}

\author{
  Ege Erdil \\
  Epoch AI \\
   \And
  Tamay Besiroglu \\
  Epoch AI, MIT FutureTech \\
}

\begin{document}

\maketitle

\begin{abstract} 
\small 
We analyze whether significant AI automation could accelerate economic growth by about an order of magnitude, similarly to how the Industrial Revolution accelerated global growth. We identify three primary drivers for such growth: 1) the scalability of an AI labor force restoring a regime of increasing returns to scale, 2) the rapid expansion of an AI labor force, and 3) the dramatic increase in output from rapid automation occurring over a brief period. Against this backdrop, we evaluate nine counterarguments, including regulatory hurdles, production bottlenecks, alignment issues, and the pace of automation. We argue that these counterarguments do not decisively rule out explosive growth. We conclude that AI systems capable of broadly substituting for human labor could plausibly lead to economic growth accelerating by an order of magnitude. However, high confidence in this outcome is unwarranted, given current uncertainties about the intensity of regulatory responses to AI, potential production bottlenecks from hard-to-quickly-accumulate inputs such as land, energy, and capital, the economic value of superhuman abilities, and the rate at which AI automation could occur.\blfootnote{We thank Jaime Sevilla, Tom Davidson, Carl Shulman, Anson Ho, Matthew Barnett, Neil Thompson, Sam Manning, Anton Korinek, and Amelia Michael for helpful feedback and Maria de la Lama for the illustrations. We are grateful to Open Philanthropy for support for this project.}
\end{abstract}

{
  \hypersetup{linkcolor=black}
  \tableofcontents
}

\newpage

\section{Introduction}
\label{sec:introduction}

Artificial intelligence (AI) possesses enormous potential to transform the economy by automating a large share of tasks performed by human labor. There has been growing interest in the possibility that advanced artificial intelligence (AI) systems could drive explosive economic growth, meaning growth an order of magnitude faster than current rates.

The idea of AI's potential to automate many or even all tasks presently undertaken by labor has drawn considerable interest from economists. A review by \cite{trammell2020economic} synthesizes various economic models to explore scenarios ranging from moderate growth acceleration to explosive growth, where growth rates themselves increase without bound or where output approaches infinity in finite time. These models demonstrate the theoretical possibility of AI-driven transformative growth through mechanisms such as increased capital substitutability for labor, task automation, and enhanced productivity in research and development, while also highlighting potential challenges and uncertainties in these scenarios.

An important such contribution to the literature is that of \cite{aghion2018artificial}, who model AI as the latest form of automation, potentially automating previously unreachable tasks. They show that AI capable of automating R\&D could lead to increased growth rates or even a singularity under certain conditions. However, the growth impact of AI may be limited by `Baumol effects' where growth is constrained by sectors or tasks that remain essential but resistant to productivity improvements. This insight helps reconcile increasing automation with observed patterns of balanced growth and stable factor shares. Their model demonstrates how combining automation with `Baumol effects' can generate a range of growth outcomes, including scenarios with nearly complete automation yet still-limited growth rates.

Prior work has sought to provide quantative ballpark estimates about the rates of growth that could be achieved with substantial AI automation. \cite{hanson2001economic} analyzes AI in the context of a neo-classical growth model. Hanson argues that initially, expensive AI systems only perform tasks where they have the strongest advantage over humans, but eventually, they do most jobs. Using conservative parameter estimates, he shows that a simple model suggests a rapid transition: before the widespread use of AI robotics, output grows at a familiar 4.3\% per year, but after machine intelligence, the growth rate increases to be on the order of the rate at which computing capacity improves (40\% per year in \cite{hanson2001economic}).

\begin{wrapfigure}{r}{0.35\textwidth}
    \centering
    \includegraphics[width=0.35\textwidth]{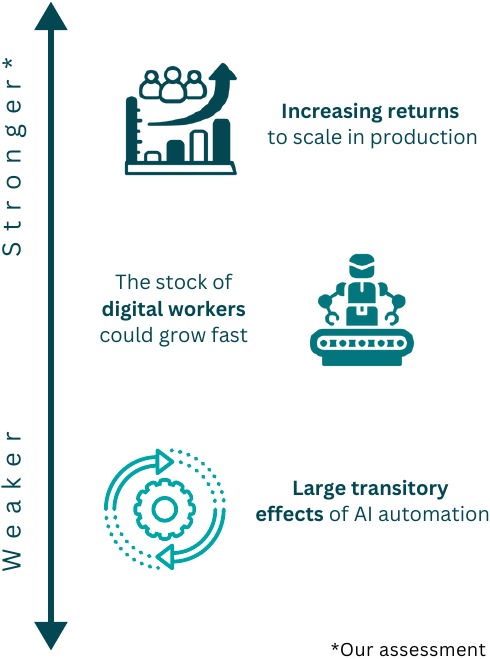}
    \label{fig:fig1}
\end{wrapfigure}
More recently, \cite{davidson2021could} examines the potential for advanced artificial intelligence (AI) to drive explosive economic growth, defined as annual growth in gross world product (GWP) exceeding 30\%. Davidson argues that if AI enables capital to effectively substitute for labor, it could rekindle the increasing returns to accumulable inputs like technology, which historically led to super-exponential growth until the demographic transition around 1900. Building on the analysis of various growth models, Davidson tentatively assigns a 30\% probability to explosive growth this century, conditional on the development of sufficiently advanced AI systems. 


We examine several key arguments for and against the possibility of explosive growth from AI—growth an order of magnitude greater than is typical in today's advanced economies. We detail these arguments and tentatively assess their strength (see Figure \ref{fig:fig1}). Our focus is on providing a quantitative foundation for key considerations, such as potential bottlenecks in automation, preferences for human-produced goods and services, and technical and regulatory challenges in implementing AI systems. By analyzing these factors, we aim to quantitatively evaluate the range of growth rates that might be achievable through AI automation, taking into account the constraints imposed by previously identified limiting factors.

We present three arguments supporting the possibility of explosive growth from AI, all based on the concept that AI could serve as an accumulable substitute for human labor. The first argument posits that AI could enhance technological progress, leading to increasing returns to scale and accelerating productivity gains. The second suggests that AI could dramatically expand the total effective workforce by combining human and artificial workers. The third focuses on the potential for rapid automation to cause a significant, though possibly temporary, surge in economic output. These arguments explore different mechanisms by which AI might drive extraordinary economic growth rates.

\newpage 

Overall, our assessment highlights four themes:

\paragraph{Growth theory models generally predict explosive growth} Standard models of economic growth consistently predict explosive growth when AI can effectively substitute for human labor across a majority or all economic tasks. This prediction holds true across a wide range of model types, including semi-endogenous and exogenous growth models, those with increasing or constant returns to scale, and models that incorporate investment delays. The robustness of these predictions is contingent on the use of realistic parameter values derived from empirical economic data and plausible projections of AI capabilities.

\paragraph{Regulation could restrict AI-driven growth} Regulation could restrict AI-driven growth by limiting AI development and deployment, potentially preventing order-of-magnitude increases in economic growth. However, such paths generally require lasting global coordination and potentially exerting control over many distributed actors, which might be infeasible given both the strengths of relevant incentives to develop and deploy advanced AI and the falling costs of AI training stemming from algorithmic and hardware technology advances.

\paragraph{Many arguments against explosive growth lack quantitative specificity or are otherwise weak} There are numerous arguments against explosive growth from AI that falter in providing quantitative specifics. For instance, some posit that fundamental physical limits or non-accumulable factors of production will rapidly bottleneck growth post AI automation, yet they fall short in quantitatively bounding the growth accelerations permitted by such constraints in a compelling manner. Other objections, such as that humans might strongly disprefer consuming AI-produced goods or services, may also fail to thoroughly take seriously genuine AI that is actually able to flexibly substituting for human labor across a wide range of tasks.

\paragraph{It is difficult to rule out explosive growth from AI, but that this should happen is far from certain} We think that the odds of widespread automation and subsequent explosive growth by the end of this century are about even. However, we caution against high confidence in this estimate for two reasons. First, there are several credible counterarguments that, while not decisive, raise legitimate reservations. Second, predicting explosive growth requires extrapolating economic models beyond their empirically validated domains, introducing significant uncertainty.

\begin{wrapfigure}{r}{0.6\textwidth}
    \centering
    \includegraphics[width=0.6\textwidth]{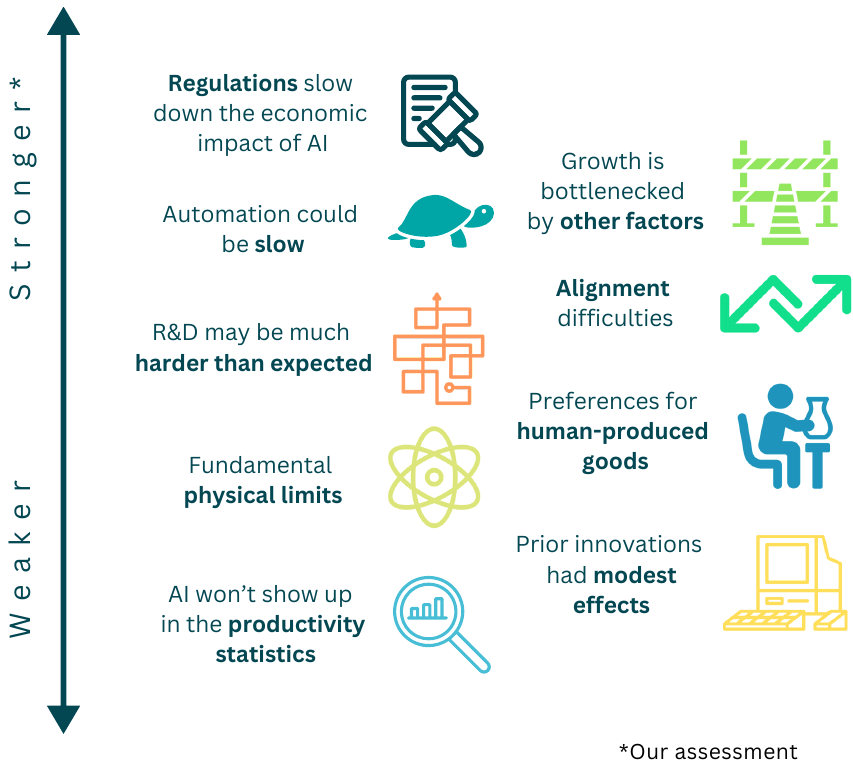}
    \label{fig:fig2}
\end{wrapfigure}

In this work, we will refer to ``explosive growth" as growth an order of magnitude greater than what is typical in today's frontier economies. Specifically, we define this as annual real gross world product (GWP) exceeding \( 130 \% \) of its maximum value over all previous years. This definition is consistent with prior definitions (e.g. \cite{davidson2021could}), and it precludes scenarios in which the level of GWP crashes (due to, e.g., some disaster) and then recovers quickly. By `economic output,' we refer to measured output figures produced by relevant statistical agencies under conditions at least as favorable as those in current frontier economies. This includes incorporating new product varieties, using adequate sampling intervals, and other standard practices.

We analyze a dozen arguments for and against explosive growth from AI capable of substantially automating economically valuable tasks. Each argument is first summarized concisely before a deeper examination aims to give a quantitative sense of how it might permit or rule out certain growth rates. After thoroughly assessing each argument, we evaluate its importance in assessing the probability of AI-induced explosive growth. To ground our quantitative estimates, appendices provide relevant economic growth models and data. We offer calibrated probability estimates for each argument being decisive in determining if explosive growth occurs. These judgments are based on a defined likelihood scale that we introduce in \hyperref[sec:appendix-a]{Appendix A}.

\section{Arguments in favor of the explosive growth hypothesis}
\label{sec:arguments-in-favor}

In this section, we present three arguments for explosive economic growth driven by the advent of AI capable of flexibly substituting for human labor. Firstly, we demonstrate that increasing returns to scale in semi-endogenous growth models generally produces explosive growth when labor is accumulable, meaning that the stock can be increased by reinvestment of output. Secondly, we extend our analysis to exogenous growth models, and show that explosive growth could emerge even without increasing returns to scale. Thirdly, we argue substantial automation happening in a brief window in time could raise the level of output sufficiently high to give rise to explosive growth. A key mechanism underpinning all these arguments is the rapid or accelerating expansion of the total labor force, which includes AI workers alongside humans due AI automation.

\subsection{Increasing returns to scale in production gives rise to explosive growth}
\label{sec:increasing-returns-to-scale}

One argument for explosive growth from AI invokes the increasing-returns production implied by standard R\&D-based growth models. In such models, when AI suitably substitutes for human labor, all factors of production become ``accumulable" so that these can be increased through investment. Notably, this gives rise to a feedback mechanism where greater output gives rise to an increase in inputs that give rise to a greater-than-proportional increase in output. Hence, such models generically predict super-exponential growth conditional on AI that suitably substitutes for human labor. The striking feature of endogenous growth models to produce explosive growth was previously pointed out, by among others, \cite{trammell2020economic}.

Standard R\&D-based growth models with increasing returns to scale predict super-exponential growth when AI can effectively substitute for human labor. To illustrate this, consider a generalized version of an R\&D-based growth model. This model exhibits increasing returns due to the nonrivalrous nature of ideas.
\begin{equation}
     Y(A, K) = A K^\beta
\end{equation}
where \( A \) represents total factor productivity and $K$ is the stock of capital (machines, computers, etc.) Capital accumulates in line with dedicated investment, as does total factor productivity. However, total factor productivity investment have diminishing marginal returns as ideas get ``harder to find" (as is well-documented in, for example, \cite{bloom2020}). Formally:
\begin{equation}
\frac{1}{A}\frac{dA}{dt} \propto A^{-\phi} I_A^{\lambda} 
\end{equation}
Standardly, motivated by the so-called ``replication argument", we might suppose that $\beta =1$. However, this assumption is not at all needed for our conclusion. Indeed, $\beta<1$ still produces increasing returns to scale as long as the returns to idea-production diminish sufficiently slowly.

In particular, we show in \hyperref[sec:appendix-semi-endog]{Appendix B} that as long as $\lambda/\phi + \beta > 1$, the economy exhibits increasing returns, which implies that such an economy will grow hyperbolically, i.e. as described by the differential equation $\frac{dY}{dt} \sim Y^c$, where \( c \) is the returns to scale parameter (which is $>1$ whenever $\lambda/\phi + \beta > 1$).

\cite{bloom2020}, which provides us perhaps with the best estimates of the extent to which ideas get ``harder to find", estimates that \( \lambda/\phi \approx 0.32 \) for total factor productivity in the US economy. Based on this, we find that hyperbolic growth occurs with values of $\beta$ as low as $1-0.32 = 0.68$. Hence, hyperbolic growth with AI is predicted by R\&D-based growth models even if we do not strictly accept the conclusion of the replication argument that \( \beta = 1 \). This point, that avoiding increasing returns to scale is difficult to avoid even when ideas get ``harder to find" over time, has been noted elsewhere, notably by \cite{davidson2021could}. Indeed, this outcome is consistent with fairly conservative assumption on there being decreasing returns on inputs to final goods production.


We think it is worth taking this argument seriously. R\&D-based growth models, and in particular, the semi-endogenous version, offer adequate explanations of recent and distant economic history, as has been noted in the literature. As such, the fact that it robustly predicts explosive growth from AI that suitably substitutes for human labor should be considered a relatively strong argument. Although obtaining high-quality empirical evidence to decide between competing growth theories is challenging, the semi-endogenous account performs relatively well in explanaining growth under Malthusian conditions, the constancy of recent growth rates, as well as maximum long-term growth rates (see Table \ref{tab:pred_explan}).

\begin{table}[htbp]
\small
    \centering
    \begin{tabular}{p{0.2\textwidth}p{0.75\textwidth}}
        \toprule
        \textbf{Prediction} & \textbf{Explanation} \\
        \midrule
        Economic growth acceleration under Malthusian conditions & An acceleration of economic growth when the size of the population is limited by the available technology (see also \cite{kremer1993population}). This prediction is in line with the observed acceleration over recorded economic history, such as that of \cite{bolt2020maddison}. While there is reasonable debate about how closely models that predict gradual economic acceleration fit distant economic data, and the extent to which such data is reliable (see, e.g. \cite{garfinkel2020economic, roodman2020probability}), this model arguably captures key dynamics of the data. \\
        \midrule
        Non-increasing growth in global output & Non-increasing growth in global output in the mid-20th century concurrent with the observed slowing rates of population growth in middle and high-income countries (\cite{jones2022past}). The model predicts that slowing rates of population growth produce slowing rates of output growth, all else equal. It therefore does a decent job accounting for the general pattern of 20th-century growth. There is furthermore evidence that the semi-endogenous growth model fits recent empirical data on output, multifactor productivity, research intensity better than other models (see, e.g. \cite{kruse2017testing, herzer2022semi}). \\
        \midrule
        Maximum observed rate of long-term economic growth & The maximum observed rate of long-term economic growth should be on the order of the maximum rate of population growth.\footnote{Assuming some dimensionless parameters such as the labor elasticity of output are close to \( 1 \), which in practice they appear to be.} Semi-endogenous growth theory predicts that growth in output should be close to the rates of population growth (see \hyperref[sec:appendix-bounds]{Appendix C}), and should therefore be on the order of 3\% per year, which is consistent with historical data. \\
        \bottomrule
    \end{tabular}
    \caption{\small \centering Summary of key predictions from the semi-endogenous growth theory with corresponding explanations and references.}
    \label{tab:pred_explan}
\end{table}

Nevertheless, it might still be appropriate to put some weight on alternative explanations, such as accounts of economic growth that consider institutions as more fundamental (e.g. \cite{north1990institutions, acemoglu2005institutions}) and appropriately be less confident that simply scaling the labor force will lead to explosive growth. However, the basic picture here still seems persuasive to us even if for some reason we believe this is not a good account of what happened in economic history, and so we still think this argument is strong in the absence of more specific critiques.


Overall, semi-endogenous growth theory offers a simple framework for understanding historical trends and patterns of economic growth. Although obtaining high-quality empirical evidence on growth theories remains challenging, the semi-endogenous account predicting explosive growth from AI systems that provide suitable substitutes for human labor presents moderately strong evidence supporting the explosive growth hypothesis.

\subsection{The stock of digital workers could grow fast}
\label{sec:argument-computing-costs}

The stock of AI systems that substitute for human workers could grow very fast once such systems have become technically feasible, which by itself could potentially expand output massively. Relaxing our earlier assumption of increasing returns to scale, we can show that even a simple exogenous growth model predicts explosive growth from AI because the stock of AI systems performing tasks that human labor previously did could grow sufficiently rapidly. Consider an exogenous growth model without technological progress:
\begin{equation}
    Y(t) = A L(t)^{\alpha}  K(t)^{1-\alpha}, 
\end{equation}
Here $L$ refers to workers: either digital workers in the form of AI systems or human workers. With the development of AI that presents a suitable substitute for human labor, we can suppose that the stocks of labor and capital grow as a result of investment:
\begin{equation}
    \frac{dL(t)}{dt} = sfY(t)/\bar{c} - \delta_L L, \,
    \frac{dK(t)}{dt} = s(1-f)Y(t) - \delta_K K, 
\end{equation}
where $f$ is the fraction of investment channelled towards AI, $ s $ is the saving rate of the economy. $\bar{c}$ denote the average dollar-costs (on compute and electricity) of building an AI system that performs the same amount of work as a human laborer. \( \delta_L, \delta_K \) are the depreciation rates for the effective labor and capital stocks, respectively. Assuming that $A$ is constant, some algebra (see \hyperref[appendix-digital]{Appendix D}) combined with the parametric assumptions presented in Table \ref{tab:parameters} reveals that the steady-state rate of growth in this model exceeds 30\% per year if
\begin{equation}
    \bar{c} \leq s^{10/7} \cdot 150,000 \, \$/\text{worker}.
\end{equation}

\begin{table}[htbp]
    \centering
    \begin{tabular}{@{}p{4cm}p{3cm}@{}}
        \toprule
        \centering\textbf{Parameter} & \centering\textbf{Value} \tabularnewline
        \midrule
        \centering Value of US capital stock & \centering \$70T \tabularnewline
        \centering US Labor Force & \centering 165M \tabularnewline
        \centering $\alpha$ & \centering 0.7 \tabularnewline
        \centering $\delta_{L}, \delta_{K}$ & \centering $\ll$ 30\% \tabularnewline
        \bottomrule
    \end{tabular}
    \captionsetup{width=0.75\linewidth} 
    \caption{\small Summary of key parameters and their values. \(\delta_{L}, \delta_{K}\) denote the depreciation rates for effective labor and capital stocks, respectively.}
    \label{tab:parameters}
\end{table}

Hence if the cost of running an AI that substitutes for a human worker (\( \bar{c} \) whose units are \( \$/\text{worker} \)) is sufficiently low, exogenous growth models predict that the effective labor stock should grow sufficiently fast to give rise to explosive growth.\footnote{A similar argument to the effect that a rapidly expanding ``digital workforce" can result in massive expansions in output has previously been made by \cite{Karnofsky2021}.}

We can provide a rough estimate of AI runtime costs by relying on estimates of the cost of computation and the estimated cost of running the human brain. Right now, machine learning hardware costs around \( 2 \times 10^{18} \, \text{FLOP}/(\$ \cdot \text{year}) \),\footnote{The current flagship datacenter GPU, the H100, can perform around 4000 TeraFLOP/s in 8-bit precision and costs around \$30,000. Conservatively assuming this is ran at 40\% utilization for three years with a \$30,000 energy bill, this amounts to \(\approx 2 \mathrm{e}18 \, \text{FLOP}/(\$ \cdot \text{year})\).} and \cite{carlsmith2020} provides a best-guess estimate of \( 10^{15} \, \text{FLOP}/\text{s} \approx 3 \times 10^{22} \, \text{FLOP}/\text{year} \) for the rate of computation done by the human brain. Combining these two estimates suggests a value of around \( \bar{c} = 1.5 \times 10^4 \, \$/\text{worker} \). In our model, this is consistent with explosive growth if
\begin{align}
    (1.5 \times 10^4) &\leq s^{10/7} \cdot (1.5 \times 10^5) \\
    0.1 &\leq s^{10/7} \\
    0.2 &\leq s.
\end{align}
In other words, this would hold if savings rates are in line with saving rates that have been historically observed in Western countries and significantly lower than saving rates that have been observed in East Asian countries such as Japan, China, and Singapore. In addition, saving could be higher under AI-driven growth, given that AI could increase the productivity of capital investments, and result in concentrating wealth to those with a high propensity to save (\cite{trammell2020economic}), though it could also lower savings due to consumption smoothing considerations.\footnote{We assume that the depreciation rates of the stock of compute and capital ($\delta_K$ and $\delta_L$) are assumed to be neglible compared to the growth rate (see Table \ref{tab:parameters}). If we were to relax this assumption then we need precise estimates of these numbers and in general need higher saving rates. However, even with depreciation rates \(\sim 30 \%/\text{year} \), we only need to double the savings rate to \( s = 0.4 \) to still get explosive growth, which has historically been observed in e.g. East Asian countries.}

Remarkably, our result holds even if we assume substantial delays in investment. In \hyperref[sec:appendix-robustness]{Appendix E} we show that the above argument is still sound even when ``realized investment" is an exponential moving average of past inputs to investment. That is, even if there are significant delays in increasing the stock of AI workers over a few years, the overall argument remains valid, with only a minor reduction in the predicted growth rate.

Overall, this calculation suggests that, even if we conservatively assume constant hardware prices and constant returns to scale, if AIs are as productive as the average US worker explosive growth is still a plausible outcome of labor becoming accumulable if our AI software can match the efficiency of the human brain.

A key parameter in this endogenous growth model involving AI-driven automation is  \( \bar c \), the average dollar-costs of building an AI system capable of flexibly substituting for human labor. It is worth noting the simplifying assumptions about the cost of AI in the preceding analysis.

The model's conclusions rest heavily on estimates of the computational requirements of the human brain, which are marked by considerable uncertainty. If we were to consider the higher-end estimates of the computation costs of running the human brain in \cite{carlsmith2020} of $1 \mathrm{e}16 \flops$, explosive growth looks unlikely with current prices.

However, it is important to note that hardware prices are expected to decrease considerably over time, with a current halving time of roughly 2.5 years (\cite{epoch2022trendsingpupriceperformance}). This indicates that the cost of running a human-equivalent AI is likely to become more affordable in the future. Therefore, the argument presented in the analysis becomes more persuasive if one anticipates that sufficiently capable AI will take around 10 to 20 years to develop, a period during which computer hardware could become one or two orders of magnitude more cost-effective. This dynamic could potentially amplify the economic growth impact of labor substitution by AI. In addition to this, it is also plausible that \( \bar c \) is lower because AIs could be more capable than humans at runtime compute parity. Here are a few reasons why we might expect this to be the case:

\begin{enumerate}
    \item A single AI system trained only once can be deployed in many different settings in the economy given a sufficient runtime compute budget, while this is impossible to do for humans. In other words, it is much easier to copy AI systems than it is to copy humans.

    This has many beneficial effects. It allows us to amortize the cost of training large systems over a vast number of runtime instances, something impossible to do with human lifetime learning. In addition, it means we can pick the best-performing systems at a given runtime compute level and simply copy those, instead of sampling from a wide distribution of conscientiousness, intelligence, communication skills, etc. that we must do when the labor force is made up of humans.

    \item Software progress on AI capabilities might not stop at human levels. Indeed, there is no particularly good reason to suppose that human brains are optimal from the point of view of converting runtime compute into capabilities, given that humans are evidence that previous species were not optimal. Even one or two orders of magnitude of decrease in \( \bar c \) from software progress would strengthen the argument in this section considerably.
\end{enumerate}

Further considerations include that the analysis relies on a static calculation of the costs of computation that does not account for potential price effects. In other words, it overlooks how demand could influence the price of computation. Moreover, we do not account for the cost of robotic systems in addition to the computational costs of running the software. While state-of-the-art industrial robotic systems are currently, for e.g. spot welding, are on the order of \$100k per unit (\cite{sirkin2015robotics}), it is difficult to predict how much this would add to the cost basis, \( \bar c \). This is because there could be substantial reductions in prices as we proceed along a learning curve as robotics usage expands (\cite{korus2019}).

An important criticism of this argument is that scaling output along the \textit{intensive margin} and the \textit{extensive margin} might be meaningfully different. For instance, it might very well be true that doubling the world population over a sufficiently long period of time leads to a doubling in gross world product, but without this increase in population leading to faster technological progress, per capita income would stay the same. If we do not count the consumption of AIs as part of GWP in our model, then our thesis is that increasing the number of AIs will lead to higher per capita consumption among humans, and perhaps it is more difficult to get explosive growth this way without being able to scale the quality of the services in the economy.

We think there is some kernel of truth in this argument, and we expect it to make explosive growth significantly more difficult in worlds where AI-driven automation is unable to meaningfully accelerate R\&D, but some scaling along the intensive margin is possible even without technological advances. There are already substantial differences in personal income across the world, and even within rich countries. In most countries, simply raising the average standard of living in the country to the standards enjoyed by the wealthiest residents would lead to orders of magnitude increase in gross domestic product, and we know that if resource constraints are sufficiently loose, doing this requires no new technology. Resource constraints could of course pose obstacles, but those are no more binding when we're talking about an increase along the intensive margin than they are when the increase happens along the extensive margin instead.


Summing up, even without the assumption of increasing returns to scale, standard economic growth models predict substantial acceleration in economic growth rates if we assume substitutes for human labor at realistic costs in the model. While we do not strongly endorse the conclusions of this calculation due to the many simplifications we make throughout, we think the argument still provides evidence that explosive growth is not unlikely, as it occurs even in the absence of endogenous technological progress or in the absence of substantial improvements in the cost of computation.

\subsection{AI automation could have massive transitory effects}
\label{sec:transitory-effects}

In growth theory, there is an qualitative distinction between \textit{growth effects} and \textit{level effects} (\cite{lucas1988mechanics}). A growth effect is assumed to be either permanent or last for a long time (e.g. changes to the steady state or balanced growth path), while a level effect is a one-time, transitory increase in the level of economic output that does not translate into higher growth in the future.

It might be the case that even if AI fails to lead to a long-term \textit{growth effect}, there might still be a \textit{level effect} from human-level artificial intelligence being deployed throughout the economy, and a change in the level of gross world product that happens over a sufficiently short window of time could lead to transitory growth rates that clear the threshold of ``explosive growth".

To quantify these effects, consider a toy model in which output is produced by a CES production function over a unit continuum of tasks:
\begin{equation}
 Y = A \left( \int_0^1 I_i^{\rho} \, di \right)^{1/\rho} 
\end{equation}
where \( A > 0 \) is a measure of productivity. We will not be explicit about what the inputs \( I_i \) represent for the sake of generality, but we assume that there is some total stock of inputs \( I \) available in the economy that can be allocated across different tasks. Moreover, $\rho<0$ so that tasks are complements, thereby giving rise to `bottlenecks' in production: the larger negative values of $\rho$, the more severe the bottlenecks.

Let $f$ denote the fraction of tasks that \textit{cannot} be cheaply automated. We show that when $f$ is relatively large (e.g. 10\%), the level effect from AI automation is very substantial, even despite substantial bottlenecks in production.
\begin{figure}[h!]
    \centering
    \includegraphics[width=0.65\textwidth]{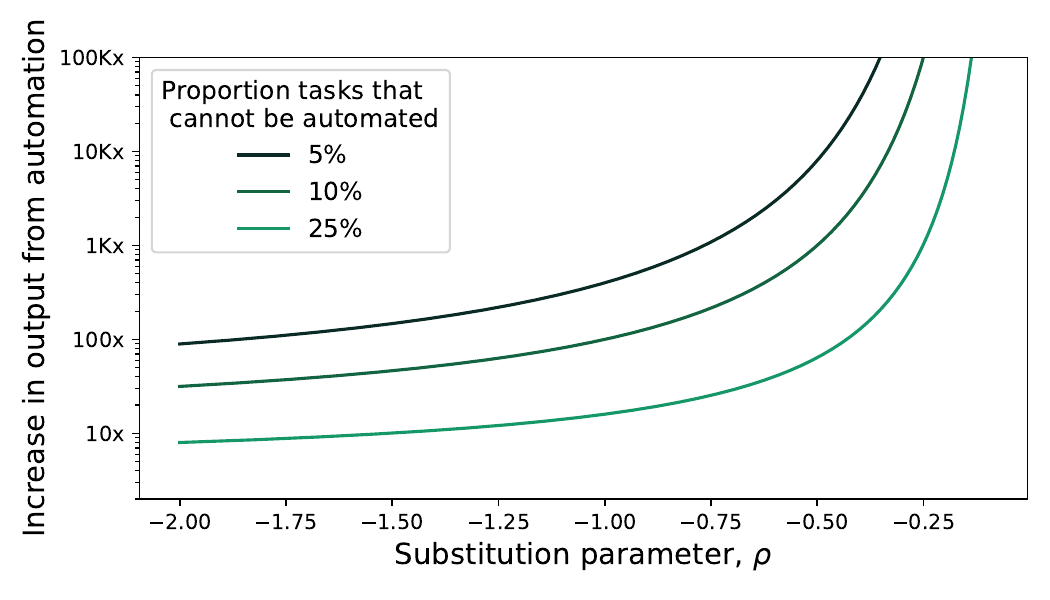}
    \caption{\small \centering Level effects of partial AI automation for various different values the substitution parameter \(\rho\) of the CES aggregator function.}
    \label{fig:transitory}
\end{figure}
Given that there is some total stock of inputs \( I \) available in the economy that can be allocated across different tasks, we have the constraint: 
\begin{equation}
 \int_0^1 I_i \, di = I, \, \, \, \forall i \, I_i \geq 0 
\end{equation}
When \( \rho < 0 \) so that the tasks are complements, \( Y \) is optimized when \( I_i = I \) for all \( i \) and therefore \( Y = AI \).

To see the impact of automation, suppose that for some \( 0 \leq f \leq 1 \), a fraction \( 1-f \) of tasks are ``cheaply automated". In practice, this is likely to mean that we get many orders of magnitude more inputs on these tasks after the automation than before. When complementarity effects are strong (so when \( \rho \ll 0 \)) we can approximate by assuming infinite input on automated tasks instead, as this simplifies the calculation without making much of a difference to the final result. In this case, our optimization problem becomes
\begin{equation}
Y = A \left( \int_0^f I_i^{\rho} \, di \right)^{1/\rho}, \hspace{0.1cm}\text{subject to} \hspace{0.1cm} \int_0^f I_i \, di = I, \, \, \, \forall i \, I_i \geq 0.
\end{equation}
This problem, as before, is solved by setting the inputs of all tasks equal to each other: \( I_i = I/f \) for all \( i \). In this case, we get that \( Y = AI f^{(1-\rho)/\rho} \), so GWP is higher by a factor of \( f^{(1-\rho)/\rho} \). Figure \ref{fig:transitory} contains the values of this function evaluated at some plausible values of \( f \) and \( \rho \).

For example, \cite{knoblach2020elasticity} looks at the elasticity of substitution between capital and labor in the US economy and finds a plausible range from \( 0.45 \) to \( 0.87 \), which corresponds to values of \( \rho = (\sigma-1)/\sigma \) ranging from \( -1.2 \) to \( -0.14 \). If $f=0.1$ and $\rho = -1$, then the proportional increase in output is 100-fold. Such a large level-effect is likely to produce explosive growth unless it is spread out over a time frame much longer than a decade decades. 

By contrast, parameter values that would block massive increases in output are on the order of \( \rho < -2 \), which are perhaps below the standard range that is considered plausible, implying stronger complementarities between tasks than we currently believe exist between capital and labor. 


It is worth reviewing the potential weaknesses in our argument for explosive growth through AI automation. This argument may not hold for several reasons:

\begin{enumerate}
    \item The approximation that AI will be infinitely productive on automated tasks makes the numbers look more impressive than they should when \( \rho, f \) are close to zero. For instance, if we believe AI will only ever contribute nine times the input intensity on any automated task, the most we can get out of AI automation is an order of magnitude increase in gross world product, even assuming full automation.

    Indeed, if we rely on our estimates of the computational cost of running the human brain from \hyperref[sec:argument-computing-costs]{Section 2.2.}, we can estimate that all the computing hardware available in the world today can perhaps run 100 million simulated workers at most. If that is the best we can do, human-level AI software will fall far short of increasing input intensities at all, let alone setting them to infinity. In the future, this can be overcome by manufacturing more chips, improving hardware efficiency, etc. but these are out of the scope of this argument.

    \item We do not have good information about what value for \( f \) should be considered plausible. In a world where full automation is attained, we can turn our model into a coarse approximation and interpret \( f \) as the fraction of tasks where running human-equivalent AIs is comparable to or more expensive than employing humans to perform the same tasks, but there is no obvious reason within the framework of this argument why this quantity should be e.g. less than \( 25 \% \). Our previous calculations based on the cost of human brains are suggestive that this number should be small, but conditional on the increasing returns to scale and digital worker cost arguments failing, we might not want to put substantial weight on this argument either.

    \item Even if the argument gives us correct values for the factor increase in GWP we should expect, this increase can simply drag out over a sufficiently long period of time such that we do not get an explosive rate of growth at any point. We address some objections which argue for such a possibility \hyperref[sec:why-not-explosive-growth]{Section 3}, but our responses are not decisive and we have to concede that long delays are indeed possible.
\end{enumerate}



This argument indicates that explosive growth remains plausible even without full AI automation and with humans still occupying critical economic roles that could potentially bottleneck production. As such, it serves as a `worst-case scenario' argument, suggesting some probability of explosive growth even under these less favorable conditions. However, given the limitations we outlined above, we consider this argument less robust than our other cases for explosive growth. We advise caution in relying on this argument, particularly in scenarios where our stronger arguments for explosive growth do not hold.

\section{Arguments against the explosive growth hypothesis}
\label{sec:why-not-explosive-growth}

In this section, we provide accounts of arguments against AI-driven explosive growth. For each, we assess their plausibility and, where possible, attempt to estimate the permitted growth rates the argument implies. While several of the arguments initially seem concerning, upon closer analysis most do not appear decisive. However, a few remain non-trivial objections that could plausibly reduce the probability of explosive growth, especially in conjunction. We examine each argument in turn and aim to draw tentative conclusions about their effects on the likelihood of explosive growth.

\subsection{Regulations can slow down the economic impact of AI}


This objection states that the training or deployment of AI systems will be sufficiently impeded by regulation to reduce the economic growth effects of AI. The possibility of the growth effects from AI automation being curtailed by regulation features, for example, in \cite{jones2021review}, \cite{Yudkowsky2021TakeoffSpeeds} and \cite{garfinekl2021review}. Presumably, there are many reasons regulatory restrictions might arise: generic apprehension regarding powerful new technologies, concerns about privacy or intellectual property leading to a shortage of training data, unwillingness to let AI systems perform tasks that can be automated without human supervision due to concerns about legal liabilities, etc. These regulations may well be prudent, potentially prioritizing safety and social welfare over economic growth. However, they could also prevent the realization of explosive growth that AI might otherwise produce. The core argument is that even if AI has the potential to generate explosive growth, coordinated efforts by governments or international bodies to slow this process could prevent this outcome from materializing.


Within the deep learning paradigm that has been dominant in AI research over the past decade, what seems to matter most for the performance of AI systems are the number of examples they see during training and the number of parameters they have -- which is in turn dictated by the amount of compute developers have at their disposal. Quantitative support for this statement is provided by the growing literature on \textit{scaling laws}, which describe the performance of a deep learning model in terms of a few macroscopic properties of the model such as the parameter count and the training dataset size. For more on this in the context of large language models, see \cite{kaplan2020scaling} and \cite{hoffmann2022training}.

In light of this, the regulation objection looks more plausible than it did a decade ago because it seems that AI development will be largely driven by access to vast amounts of data and computation. The large physical footprint of the computation capacity required for training and deploying advanced AI would likely make the process easier to regulate, and intellectual property laws can be a significant impediment to the data scaling part of the equation if they were to be interpreted in a manner unfavorable to AI labs. So we do not think we can rule out this scenario as it stands, especially if in the future there are large and visible alignment failures of AI systems that scare people into action.

However, there are effects pushing in the opposite direction. Insofar as being in possession of better AI systems becomes a matter of national security, we can expect any coordination by governments across the world to slow down AI development to be imperfect. Furthermore, the scale of the potential economic value that AI capable of widely substituting for human labor can create is enormous: it is orders of magnitude beyond any other recent innovation we can think of, mainly because of its credible potential to restore the historical trajectory of accelerating growth. These factors create strong incentives for governments to allow the widespread deployment of AI systems.

We also have to consider algorithmic progress and improving hardware efficiency. While scaling laws give a good description of the performance of ML systems at a particular level of algorithmic efficiency, over time we develop better software and this means we need fewer resources to achieve the same level of performance. \cite{hernandez2020measuring} estimates the pace of algorithmic efficiency improvements in computer vision as one doubling every 16 months and \cite{erdil2022algorithmic} estimates one doubling every 9 months, though with wide confidence intervals. If these rates of progress are at least within the right ballpark and hold up across many orders of magnitude of progress, eventually AI systems capable of widely substituting for human labor could become quite cheap to train.

In addition, the falling price of computation over time due to hardware efficiency progress means this represents an increasingly smaller fraction of global spending on computation. To keep up with these two effects, increasingly strict regimes of surveillance could eventually be required. The theoretical lower bound on the resource needs of AI set by the human brain should loom large in our thoughts here: the existence of the human brain means that in principle we do not need more energy or data than is used by a human to achieve human-level performance, and tracking every human born in the world would require a surveillance regime the likes of which we have never seen so far. We think the first AI systems capable of substituting for humans will require substantially more computation and data than the human brain does, but over time there is no reason why these costs should not fall to the level of the human brain or even further below.

In light of the above discussion, we think our baseline scenario here for AI regulation should be more like nuclear arms control and less like the regulation of nuclear energy: coordination on nuclear arms control does happen, but it is quite imperfect and hasn't stopped nuclear proliferation from taking place. This is because we think the incentives for AI adoption are more similar to the incentives for nuclear proliferation than the incentives for using nuclear energy, as the economic value that would be unlocked by AI is far greater and this also has the potential to directly translate into overwhelming military advantage against adversaries.

Here are some concrete ways in which regulation could be used to slow down the economic impact of AI:

\begin{enumerate}
    \item Place restrictions or otherwise impose additional costs on large training runs, similar to the restrictions that now exist on nuclear power. The large resource footprint of training runs past the \( 10^{27} \flop \) scale or so should make these enforceable for some time.

    \item Prohibit the use of AI for certain economic activities. For instance, laws could be created or interpreted to bar the use of AI in courtrooms or at hospitals without adequate human supervision. This would introduce an artificial bottleneck that would stop AI from fully automating some tasks.

    \item Use intellectual property laws to prevent the use of certain kinds of data for the training of AI systems. A sufficiently expansive interpretation of existing intellectual property legislation could prevent AI from being usefully monetized, reducing the incentive for private actors to invest resources into developing better AI systems.
\end{enumerate}

While implementing such regulations may hinder the development or deployment of AI, the feasibility of enacting and enforcing them remains uncertain. Firstly, it is unclear whether such policies can reliably remain enforced over a sufficiently large, possibly global, jurisdiction for multiple decades or longer. The potential value of AI deployment could be immense, with the prospect of increasing output by several orders of magnitude. Consequently, this would likely create formidable disincentives for imposing restrictions, as well as powerful incentives for eliminating or bypassing any existing constraints. Secondly, the difficulties with enforcing such restrictions might become large as software improvements bring the capital costs of AI training down. Over time, enforcing such restrictions will require increasingly ubiquitous global surveillance.

The historical record of regulating technologies that could boost output tenfold is sparse because few, if any, such technologies have previously been developed. Perhaps the closest possible analogs are a cluster of agricultural technologies that were introduced during the Neolithic Revolution or the manufacturing technologies that contributed to the Industrial Revolution (the steam engine, the spinning jenny, cotton gin). While England attempted to forestall the diffusion of some key Industrial Revolution technologies by prohibiting the emigration of skilled workers and the export of machinery, these protectionist policies proved largely ineffective (\cite{jeremy1977damming}). From the 1780s to 1840s, skilled workers, machines, and blueprints were frequently smuggled out of the country despite the bans, and by the 1840s, with industrialization advancing rapidly, the policies were seen as futile and repealed (Ibid.). In summary, England failed to meaningfully slow the international diffusion of its industrial technologies. The experience highlights the challenges of restricting technologies that offer major economic gains.

As far as we can tell, there is no compelling evidence to suggest that technologies involved in prior shocks to production technologies could have been effectively regulated with the effect of not just delaying such shocks but also substantially dampening their growth effects.


Overall, we conclude that regulating the training and deployment of AI may delay its economic impact, but there is no compelling reason to be confident that its development and application would be sufficiently prolonged to maintain historical economic growth rates for an extended period of several decades. We do not rule out the possibility, but we would judge it to be \textbf{unlikely} that regulation of the training and deployment of AI will block explosive growth.

\subsection{Output is bottlenecked by other non-accumulable factors of production}
\label{sec:other-bottlenecks}


The endogenous growth theory argument for explosive growth from the \hyperref[sec:increasing-returns-to-scale]{Section 2.1} only implies that we should expect constant returns to scale on all physically embodied inputs jointly. Labor and capital are physically embodied inputs, but they might not be the only important ones: other inputs such as energy or land could be just as important, and if they cannot be accumulated through better technology, perhaps this means AI-driven growth can get short-circuited by its dependence on these non-accumulable factors before reaching the threshold of ``explosive growth". In addition, just like population, there might be intrinsic timescales that block currently accumulable inputs such as physical capital from being accumulated arbitrarily quickly. If true, this could be a strong objection against the explosive growth view.


Some version of this argument is certainly sound: there must eventually be some resource constraints that prevent output from growing arbitrarily large. The important question about this argument is not whether it holds \textit{eventually}, but whether it holds quickly enough to preclude explosive growth.

We estimate that the diminishing returns structure on idea production implied by \cite{bloom2020} means that we need the returns to scale on accumulable inputs to be at least around \( d \approx 0.68 \) for explosive growth to occur (see \hyperref[sec:appendix-semi-endog]{Appendix B}). Theoretically, we have reason to believe that \( d = 1 \) (i.e. we have constant returns to scale) if we consider \textit{all} physically embodied inputs. However, not all such inputs may be accumulable: as a naive example, if empty space becomes a valuable resource, then regardless of how much output we invest the speed at which we can grow our access to space might be bounded by the speed of light. There is no \textit{a priori} argument which can settle the question of the returns to scale on accumulable inputs, and we must also consider the possibility that there might be strong complementarity between presently accumulable inputs such as capital and non-accumulable inputs such as land or empty space. We must examine the argument in greater detail to make a judgment about its strength.

The outside view consideration is that economic growth has accelerated by many orders of magnitude in the past: indeed, this is the empirical regularity for which the semi-endogenous growth theory provides an explanation. Factors which bottleneck this acceleration do not seem commonplace. We might look at the \( 1.5 \) order of magnitude increase in growth rates since the agricultural era as evidence that new bottlenecking factors such as population growth appear at a rate of roughly once every \( 1.5 \) orders of magnitude of acceleration, suggesting a probability of \( \sim 1 - 1/(1 + 1/1.5) = 40\% \) using the time-invariant version of Laplace's rule from \cite{epoch2022atimeinvariantversionoflaplacesrule} that one such bottleneck appears before world economic growth accelerates by one order of magnitude.  

When we get down to specifics, the most plausible bottlenecking factors that we can think of are land, energy and capital. On the energy front; on average, $4.4\mathrm{e}16$W hit the Earth (\cite{nasa2005balance}), while global yearly energy consumption is about $4\mathrm{e}13$W (\cite{owidenergy}), suggesting that energy consumption could expand by 3 orders of magnitude. Similarly, only around 1.5m km$^2$ out of the Earth's 100m km$^2$ of habitable land is urban and built-up land, which suggests that there are around 2 orders of magnitude of land that could be urbanized or built-up. Even if these are strong constraints that cannot be overcome, considering just these constraints, we have at least \( 2 \) orders of magnitude of room to scale up gross world product. If we assume no improvements in efficiency, so that resource consumption needs to be scaled up proportionally to output, such constraints would still permit explosive growth if the transition to full automation took 20 years or less. Clearly, then such constraints do not block growth accelerations at least when AI automation occurs swiftly. 

We examine the prospect that some form of physical capital could end up being a bottlenecking factor quantitatively and come to the conclusion that for the argument to block explosive growth, we need adjustments in investment to be significantly slower than the growth rate of the broader economy (see \hyperref[sec:appendix-robustness]{Appendix E}). In particular, if we can double the worldwide stock of physical capital at a rate of \( 30 \%/\text{year} \), there is no reason to suppose that explosive growth would be prevented due to investment delays or adjustment costs.

The assessment of the likelihood of physical capital becoming a bottlenecking factor, therefore, comes down to the quantitative question of how long we can expect fundamental delays to investment to be. The experience of Chinese catch-up growth shows that sustained growth rates on the order of \( 10 \%/\text{year} \) and one-time growth rates on the order of \( 15 \%/\text{year} \) have precedent in economic history. To reach the threshold of explosive growth, we only need a doubling of this final rate of increase in a world where AI will also be capable of assisting with the process of capital stock adjustment, which does not seem sufficiently far out of distribution for us to seriously doubt its feasibility.

Overall, the inside view here seems somewhat ambiguous and it is difficult to know in which direction we should update given the above paragraph. The fact that the joint returns to scale on labor and capital right now seem to be well over the threshold of \( d = 0.68 \) required for explosive growth is reasonably good evidence that we should expect at least some period of growth acceleration after full automation, but this period might be short and it might stop before we actually reach \( 30\%/\text{year} \) in gross world product growth rates.

Still, we think the threshold of \( 0.68 \) might be really  low, much lower than where existing empirical evidence places it (e.g. \cite{kariel2022returns, basu1997returns}). In addition, even in a world where this objection is valid, the argument from accumulating digital workers from \hyperref[sec:argument-computing-costs]{Section 2.2} could still produce explosive growth for some time as a consequence of the transition from human labor to AI labor. As a consequence, our estimate of the probability that this objection blocks explosive growth is substantially smaller than the naive outside view figure of \( 40 \% \).


Our final conclusion is that this argument is plausible on the outside view and the inside view evidence makes the argument seem somewhat less compelling, though is by no means sufficient to rule it out. Our final judgment is that it is \textbf{unlikely} this objection blocks explosive growth.

\subsection{Technological progress and task automation by AI will be slow} 
\label{sec:slow-tech-progress}


This argument posits that the requirements for automating different tasks in the economy span a wide range in computation, data or both. As these resources can only be accumulated in a gradual fashion, it will take a long time to get from the point where AI starts to have a large economic impact by automating tasks that are the easiest to automate to the point where AI is able to fully automate the economy, and this long waiting period will spread out the economic impact sufficiently that we end up not observing explosive growth.

The effect of this argument is similar to the argument from regulation, but the underlying driver is different. Here, the reason is a physical property of AI systems as such, and not a property of how human civilization will react to the prospect of full automation of the world economy by AI. In both cases, however, there is some force that causes the large impact of full automation to be spread out over a long period of time, and this is what precludes explosive growth.


This objection rests on an empirical claim about the relative difficulty and resource requirements of automating different tasks in the economy, specifically that the distribution of the amount of computation, data, etc. required to use AI to automate different tasks in the economy is wide and/or fat-tailed. In other words, we need \textit{some} tasks to be easy and automated early on, and \textit{some} tasks to be very difficult and to take many orders of magnitude more resources to automate.

If this objection holds, it could indeed be why explosive growth does not occur: a 4 order of magnitude (\( 4 \oom \) hereafter) increase in gross world product spread out evenly over \( 80 \) years would not produce explosive growth, for instance. A specific plausible story here is that ``physically embodied" tasks such as general-purpose robotics will be quite difficult to automate -- solving them will require large amounts of computation, data, and researcher effort.
\begin{figure}[h!]
\centering
\includegraphics[width=0.65\textwidth]{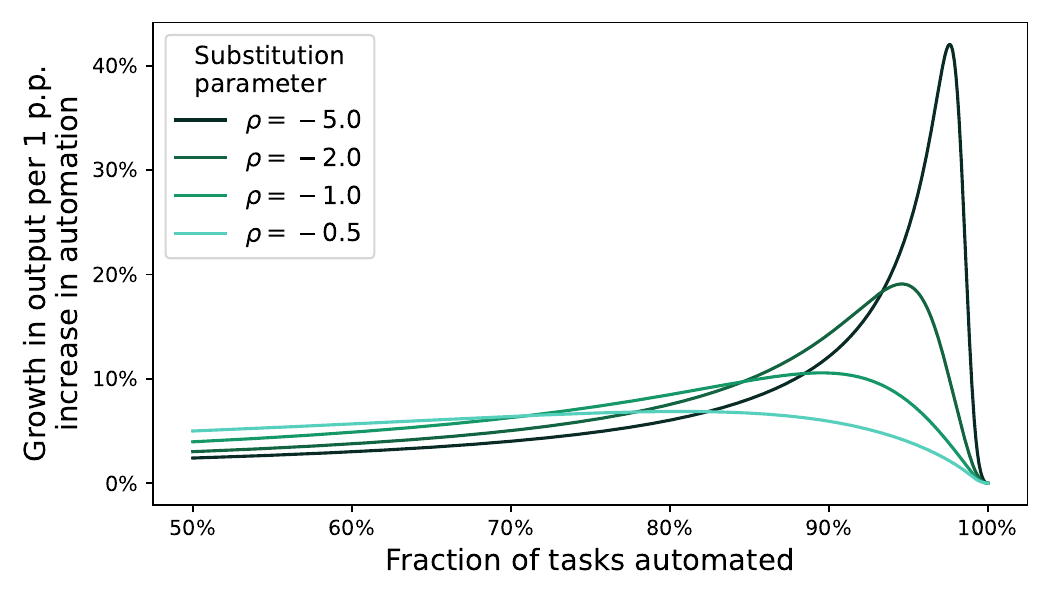}
\caption{\small \centering The distribution of growth effects from automation on gross world product across difference values of the substitution parameter $\rho$. The figure illustrates how output growth becomes increasingly back-loaded as \( \rho \) becomes more negative, indicating stronger complementarity effects. For instance, $\rho = -5$ corresponds to an elasticity of substitution of $\sigma = 1/6$ (extremely strong complementarity), and $\rho = -0.5$ corresponds to $\sigma = 2/3$ (moderately complementarity). All scenarios assume a 100x total level effect from full automation.}
\label{fig:complementarity-paradox}
\end{figure}
This is among the more compelling reasons why we might not get explosive growth. However, on the inside view, it still seems rather unlikely to be correct. There are two main reasons for this:

\begin{enumerate}
    \item Slow deployments and automation require large gaps in compute and data requirements between the point where AI starts to accelerate economic growth and the point AI is able to fully automate the world economy. However, inside-view investigations into AI (such as \cite{cotra2020forecasting,davidson2023compute}) do not usually support such large gaps.
    
    The largest plausible gap in training computation between AI starting to have a noticeable macroeconomic impact and full automation that has been suggested in such inside-view investigations is around 10 orders of magnitude, and even this gap would be crossed fairly quickly if we add up the effects of hardware scaling, improving hardware and software efficiency, etc. It is implausible that we could get a delay that is as long as 80 years, and delays on the order of 30-40 years seem like the slowest that takeoff could end up being.

    \item Even if the delay period is much longer than 40 years, in a straightforward constant elasticity of substitution (CES) world, where tasks performed in the economy are gross complements so that automated ‘outputs’ are imperfect substitutes for non-automated ‘outputs’, the final tasks to get automated are substantially more valuable than earlier tasks. This means that a constant \textit{rate} of task automation (say, automate \( 1 \% \) of tasks that humans can do in the year \( 2020 \) every year) leads to initially slow growth that becomes extremely fast towards the end, as can be seen in Figure \ref{fig:complementarity-paradox}. This intuition seems compelling: the final tasks to get automated remove the final bottlenecks in production, so if we believe full automation is actually possible it is difficult to construct a scenario in which we do not get explosive growth here.
\end{enumerate}


While we consider this objection unlikely to be correct, it is internally coherent and more compelling compared to most of the other objections. We expect it to be \textbf{unlikely} that this objection blocks explosive growth.

\subsection{Alignment difficulties could reduce the economic impact of AI}


If AI alignment—the challenge of steering artificial intelligence (AI) systems to behave according to intended goals and avoid unintended harmful behaviors—turns out to be so difficult that it is hard to get AI systems to reliably do what we want in real-world deployment, then aside from these systems being regulated more strictly, it could also simply not be in the private interest of any actor to deploy such systems at large scale. The capabilities of an AI system may seem impressive in the lab, but if private actors are unable to confidently align these systems to accomplish the tasks they want done safely, it is hard to foresee such unaligned AI generating major economic impact before these alignment problems are solved.

It might also be challenging for AI systems to be deployed to perform certain tasks without human supervision. For instance, as outlined in \cite{ji2023survey}, a common alignment problem of modern large language models is their tendency to hallucinate facts that are wrong: when asked to provide references for a claim they have made, they will often respond with references formatted according to the proper guidelines but referring to papers that do not exist. If the tendency for models to hallucinate facts cannot be entirely fixed, it might be necessary for a human in the loop to be present in any application where strict agreement with facts is highly important, which would mean there are limits to how far poorly aligned AI systems are able to automate such tasks.

There are many other paths to alignment problems leading to AI performing below the economic potential we might attribute to it based strictly on capabilities. As another example, if humans are concerned about misaligned AI systems having too much agency, they might deliberately try to engineer AI systems to be less independent in their decision-making than humans. This would then require humans to play critical decision-making roles in the economy, and then human decision-making capabilities could end up being a bottleneck in the way of explosive growth.


While the motivation behind the alignment difficulty argument is quite different from the other arguments we consider, formally its effects are likely going to be equivalent to limiting the fraction of tasks AIs are able to perform in the economy. For instance, if humans must occupy key decision-making roles in the economy, this means that effectively these tasks cannot be automated from an economic point of view. The effect of some tasks requiring human supervision to perform is similar.

This means our quantitative basis for assessing the above argument should be similar to what we outline in \hyperref[sec:transitory-effects]{Section 2.3}. If we think misalignment is likely to be so bad that e.g. \( f = 25 \% \) of tasks are likely to remain unautomated, and the elasticity of substitution across tasks \( \sigma \approx 1/3 \), then it is quite plausible that this argument blocks explosive growth. However, as discussed in the aforementioned section; \( 25 \% \) is a large fraction of the tasks in the economy, and \( \sigma \approx 1/3 \) is a high degree of complementarity across tasks. As before, we consider both of these parameter choices to be rather unfavorable, but the alignment difficulty argument pushes us to think that perhaps a \( 25 \% \) lower bound on the fraction of unautomated tasks in the economy is not as implausible as it might seem.

It is rather unclear what probability distribution is implied by this argument over the parameter \( f \) of tasks that AIs won't be able to automate ``early on". However, it seems likely to us that this distribution puts significant probability mass on values that are rather small and would block explosive growth even with moderate values of \( \sigma \), with potentially long delays in the R\&D process that would push this fraction down, echoing the arguments from the \hyperref[sec:slow-tech-progress]{Section 3.2}. Overall, our assessment is that this argument is most likely not going to block explosive growth, but its influence cannot be ruled out, especially in worlds where \( \sigma \) turns out to be smaller than we expect.


Overall, our conclusion is that alignment difficulties are \textbf{unlikely} to block explosive growth. Furthermore, this argument is in a family of arguments whose plausibility is correlated with one another due to the confounding influence of the elasticity of substitution parameter \( \sigma \), and therefore it is important to take care when aggregating the probabilities: the disjunction of these arguments is less likely than would be implied if we simply treated them as independent events and blithely multiplied their individual probabilities of not being blockers to get a final answer.

\subsection{R\&D may be harder than expected}


One argument for why we might not see explosive growth is that R\&D may be simply too hard. More precisely, the idea production function for total factor productivity (TFP) may have such unfavorable diminishing returns that it blocks the whole model from exhibiting explosive growth. As we have shown in \hyperref[sec:other-bottlenecks]{Section 3.4}, to make this work, the rate of the returns to R\&D and the returns to scale in economic production jointly need to be small enough so that the the feedback between economic inputs and output is too decoupled to give rise to accelerating growth. If we take existing estimates of the returns to R\&D in for US TFP, from \cite{bloom2020}, this argument works if the homogeneity of the production function for other non-idea outputs is no greater than \( d = 0.68 \).


We unfortunately do not have much evidence to evaluate how plausible the premise of this objection is. Although we have estimates related to $d$ for the advanced economies today, it is unclear how much these should inform us about the returns to scale in economies that are possibly bottlenecked by other factors of production. It is also appropriate to put some probability mass on the possibility that present estimates of the returns to R\&D are too aggressive, or that returns might fall over time as we make more progress in R\&D. However, even if we assume the premise that returns to R\&D are less favorable than present estimates suggest, this argument isn't sufficient to rule out explosive growth because of the argument based on the cost of computation we advance in the \hyperref[sec:argument-computing-costs]{Section 2.2}. Indeed, even in a world with exogenous technological progress and diminishing returns to scale on labor, explosive growth still remains a plausible outcome.

Because of the uncertainty about the premises of this argument and that it does not seem easy for this effect to block explosive growth even if the premises of the argument are assumed to be valid, this argument seems rather weak. We accordingly estimate a low probability that this argument is a decisive blocker. In light of the above assessment, we conclude that it is \textbf{very unlikely} that unexpected difficulties in R\&D that result in stagnating TFP growth will end up blocking explosive growth.

\subsection{AI automation will fail to show up in the productivity statistics}


Even if substantial AI automation causes explosive growth in some intuitive sense, it is possible that economic measurement will be flawed in some respects and fail to capture a possible growth acceleration. Therefore, we could end up seeing a world of rapid economic transformation in which GDP growth statistics nevertheless fall far short of the threshold of \( 30\%/\text{year} \) we set for explosive growth.


There are at least two related arguments for why substantial AI automation will fail to show up in productivity. The first is that economic output will be inaccurately measured and that this measurement error will result in a downward bias in the estimated rate of economic growth. The second related objection is that there are well-known issues with measured growth in economic output failing to capture growth in consumer surplus, so even if the measurement of output was highly reliable, estimated economic growth would fall short of growth in consumer surplus. The second objection also contends that consumer surplus is, in some sense, the more important metric.

The first argument that economic growth will be imperfectly measured and suffer attenuation bias is indeed plausible. There are many reasons why this might happen, such as:
\begin{itemize}
    \item Lag in incorporating new product varieties: Official economic agencies often fail to promptly incorporate new types of products into their metrics. For instance, the advent of electric vehicles took years to be accurately reflected in GDP calculations.
    \item Inadequate sampling intervals: Current sampling intervals may be too long to capture short bursts of rapid economic growth. 
    \item Random measurement errors: Factors like imperfect quality adjustments introduce random errors into growth estimates. Such error could introduce attenuation bias into the estimates of growth.
\end{itemize}

The first of these is in part the reason why the productivity effects in IT have been relatively meager (see, e.g. \cite{brynjolfsson1993productivity}), and the same measurement issues might similarly result in the underestimation of the effects of AI. 
On the other hand, the existing literature on the accuracy of GDP estimates suggests that these are not usually statistically biased. For example, the difference between preliminary estimates and later estimates derived from the comprehensive economic census tend not to be systematically different, at least in G7 countries (\cite{york1997reliability}) or in the US (\cite{landefeld2008taking, mankiw1986news}). Moreover, using data from six comprehensive revisions—in 2009, 2003, 1999, 1995, 1991, and 1985, \cite{fixler2011revisions} finds that the size of BEA revisions of advance GDP estimates are not correlated much at all with preliminary GDP estimates. This suggests that historical growth accelerations are not likely to be systematically underestimated, at least in the United States.

This leaves us with conflicting insights regarding the economic implications of AI. On one hand, GDP estimates in leading economies have generally proven to be unbiased and reliable. On the other hand, the economic contributions of past technological innovations like IT have been historically under-reported due to measurement issues.


However, as discussed in Sections \ref{sec:arguments-in-favor} and go on to discuss in \ref{sec:previous-tech}, the economic impact of a technology that can widely substitute for human labor could far exceed that of past technological innovations like IT. Given this, it is reasonable to expect that statistical agencies, operating under conditions at least as favorable as today's, will more accurately estimate the economic gains from AI, akin to how they track overall GDP. Relevant agencies might adapt to a faster rate of change. In an AI-automation world, agencies could face pressures to ensure that tracking and monitoring are commensurate with the pace of change. Their budgets are likely to expand in line broadly with the size of the economy, and relevant technologies used for monitoring with the sophistication of extant technology.

We think this argument is somewhat implausible, mostly because it relies strongly on the notion that output measurements will make predictable and large errors that we can anticipate but competent statistical agencies will predictably fail to address. Even with limited knowledge of these agencies' operations, we find the assumption hard to believe. 

A weaker version in which we do not claim to predict the sign of the error in advance is somewhat more convincing. In light of this objection, one's expectations of growth rates under AI automation should be more spread out. The net effect of this is depends on one's expectations of growth rates from AI automation: if one were confident in explosive growth, one should shade their probability estimates in light of additional noise. On the other hand, if one were confident that explosive growth would not occur, one should assign a greater credence to statistical agencies reporting 30\% growth rates. In the end, we consider it \textbf{unlikely} that GDP measurements will make errors sufficiently large and systematic for their measures to not show explosive growth occurring.

The second argument is based on the recognition that there are well-known issues with measured growth in economic output failing to capture growth in consumer surplus, as the former fails to capture the value of `free' IT goods, such as Wikipedia, Google search, OpenCourseWare, and so on. Perhaps, an AI-driven economy will produce a relatively larger share of goods that fail to show up on the usual output accounting. 

Existing attempts to estimate the contributions from `free' goods find the contribution of is relatively small, contributing roughly no more than one-tenth, in proportional terms, to GDP growth numbers. For example, \cite{nakamura2017measuring} estimate that including `free' content would raise U.S. GDP growth by about 0.03 percentage points per year from 1995 to 2014. Relative to the average GDP growth rate of 2.5\% over that period, this would represent a very small margin of error. Other attempts at similar accounting of the contributions from `free' content like Facebook likely find slightly larger contributions (e.g. \cite{brynjolfsson2019gdp}), but similarly suggest that this added growth  adds on the order of tens of basis points to GDP growth, at least in the United States.


In addition, even if such errors did come to pass, at some level, we do not care about productivity statistics in any fundamental sense. They are simply a useful proxy for what we wish to discuss, and if they fail to be a good proxy in the future, that does not necessarily mean our thesis about explosive growth is mistaken or that we shouldn't take action to prepare for a world in which explosive growth will occur. We find it exceptionally unlikely that this argument blocks explosive growth in a sense that we would care about, as opposed to e.g. being a measurement artifact.

\subsection{Human preferences for human-produced goods will bottleneck growth}
\label{sec:preference-bottleneck}


Humans may have a preference for human providers over AI counterparts even in economically significant service industries. Even if AI is physically capable of doing any task as well as a human can or better, there might be some tasks that are valued by humans only when they are performed by other humans. For example, we today have computer programs that can play chess better than any human player can, but top human chess players can still make money by winning tournaments. The fact that a tournament of computers would have a better quality of play is not important because part of what people want to watch is for \textit{humans} to be playing the game. 

Humans might prefer to interact with human therapists, teachers, or other service providers that involve high symbolic value and expression of identity (\cite{granulo2021preference}). Although AI systems may one day replicate some social abilities of humans, people currently tend to prefer human interaction for certain services. If such intrinsic preferences apply to a sufficient range of tasks, full automation might be impossible simply because of human preferences and not because of any physical fact about what AIs can or cannot do. This would limit gross world product as long as humans remain the ultimate consumers in the world economy and therefore the prices of goods and services are set according to their marginal utility.


This objection could in principle work assuming that all prices in the economy are set by humans, but there are two main problems with it.

\begin{enumerate}
    \item While there might be good reasons to care about what happens to gross world product, we're fundamentally more interested in questions about the ability to manipulate the physical world to get desirable outcomes. Importantly, scenarios in which AI poses a significant military risk or reshapes the physical environment around us in some substantial way can still be ``explosive" in character even if humans are setting the prices of goods and services and therefore GWP ends up being bottlenecked by human preferences of one sort or another.

    \item Even on the argument's own terms, the parameter values needed to make this story work seem quite implausible.
\end{enumerate}

The first problem is relatively straightforward, so we focus on the second problem here. Suppose that consumer utility is some monotone transformation of the CES aggregator
\begin{equation}
U = \left( \int_0^1 c_i^{\rho} \, di \right)^{1/\rho}, \quad \rho = \frac{\sigma-1}{\sigma}, \quad \rho < 0
\end{equation}
over individual consumer goods \( c_i \). If markets clear in some underlying model such that goods prices are proportional to marginal utility, GDP growth would be given by
\begin{equation}
\frac{dY}{Y} = \frac{\int_0^1 p_i d c_i \, di}{\int_0^1 p_i c_i \, di} = \frac{\int_0^1 U_i d c_i \, di}{\int_0^1 U_i c_i \, di}, \hspace{0.1cm} \text{where} \hspace{0.1cm} U_i = \frac{\partial U}{\partial c_i} = c_i^{\rho-1} U^{1-\rho},
\end{equation}
so that the expression for GDP growth simplifies to
\begin{equation}
\frac{dY}{Y} = \frac{\int_0^1 c_i^{\rho-1} d c_i \, di}{\int_0^1 c_i^{\rho} \, di} = \frac{1}{\rho} \frac{d U^{\rho}}{U^{\rho}} = \frac{dU}{U}.
\end{equation}
This equation is solved by \( Y \propto U \). Therefore, for this particular specification, GDP perfectly tracks consumer utility, and we can reason about GDP growth by using the growth of \( U \) as a proxy for it.

If a fraction \( f \) of tasks can only be done by humans by definition, and initially the \( c_i \) are all equal, then setting the output tasks that cannot be done by humans to infinity should raise \( U \) by at least a factor \( f^{1/\rho} \), and this factor would increase if we could explicitly take human labor reallocation from automated tasks to human-only tasks into account - if the technology converting human labor to output on individual tasks is constant returns to scale, for instance, then we can get this up to \( f^{(1-\rho)/\rho} \).

This is just the same expression that we dealt with in the \hyperref[sec:transitory-effects]{Section 2.3}. We present a range of parameter values to analyze the plausibility of the argument here in Table \ref{tab:scale-up-gdp}: 
\begin{table}[h]
\centering
\begin{tabular}{@{}p{3cm}p{3cm}p{3cm}p{3cm}@{}}
\toprule
\centering & \centering \textbf{\(\rho = -0.2\)} & \centering \textbf{\(\rho = -0.4\)} & \centering \textbf{\(\rho = -2\)} \tabularnewline
\midrule
\centering \textbf{\(f = 5\%\)}   & \centering \(6.4 \times 10^7\)     & \centering \(3.6 \times 10^4\)     & \centering 89 \tabularnewline
\centering \textbf{\(f = 10\%\)}  & \centering \(10^6\)                & \centering \(3.2 \times 10^3\)     & \centering 32 \tabularnewline
\centering \textbf{\(f = 25\%\)}  & \centering \(4.1 \times 10^3\)     & \centering 128                     & \centering 8  \tabularnewline
\bottomrule
\end{tabular}
\caption{\small \centering A table showing the scale-up factors we can get in GDP for various different values of the fraction of tasks that cannot be automated by AI, \( f \); and the substitution parameter \( \rho \) of the CES aggregator function.}
\label{tab:scale-up-gdp}
\end{table}

The value \( \rho = -2 \) corresponds to an elasticity of substitution \( \sigma = 1/3 \), which is conservative. Even under the pessimistic assumptions of \( f = 25 \% \) and \( \rho = -2 \), AI that broadly substitutes for human workers should produce at least \( \approx 1 \oom \) increase in gross world product. If this happens in less than a decade, it would be sufficient to produce explosive growth.

We think this scenario is pessimistic because both parameter values seem unreasonable. We think \( f = 5 \% \) to \( f = 10 \% \) are more realistic values for the  fraction of current economic tasks humans would only value if they were done by humans, and \( \sigma = 0.7 \) is a more realistic value for the elasticity of substitution in the human utility function. Combining these means we should expect around \( 3-4 \oom \) increase in GDP as a result of AI even if we accept the argument that some tasks will not get automated as a result of human preferences for those tasks to be done by humans. This is, as mentioned previously, more than enough to produce explosive growth for an extended period of time.

As a reference point, note that \( 3-4 \oom \) likely matches how much gross world product has increased since the Industrial Revolution, and plenty of this came from increased task automation. So arguments based on intrinsic human preferences for some tasks being performed by humans seem like they would have made poor predictions if we had relied upon them in the past, and accordingly, we should be skeptical of them today as well.


We think that this argument will have some effect on economic growth, but do not consider it important for three main reasons:

\begin{enumerate}
    \item It is not clear if all prices in the economy will actually be set by humans. If AIs can own property and are able to make consumption decisions as well, then gross world product would also take their preferences into account, and these preferences may not come with intrinsic demands that certain tasks must be performed by humans to be valuable.

    \item Quantitatively, the magnitude of the complementarity in the utility function and the mass of tasks that humans wish to be intrinsically done by other humans have to be quite large for this argument to block explosive growth.

    \item Even if explosive growth in gross world product is blocked, this does not necessarily mean that explosive growth is blocked in other physical variables that we might care about. These might include energy use, military strength, computer chip production, etc.
\end{enumerate}

For all of these reasons, we consider this argument to be rather weak and do not think it should lead us to update our credence in explosive growth conditional on AI downwards by a substantial amount. We consider it \textbf{very unlikely} that this argument blocks explosive growth.




\subsection{Previous technological revolutions did not lead to growth acceleration}
\label{sec:previous-tech}

We have seen many other technological innovations in the past that changed how we live our lives: computers, electricity, cars, airplanes, etc. Nevertheless, while these technologies allowed the trend growth rate of around \( 2\% \) per year per person in the US and other developed economies to continue, they didn't lead to any noticeable growth acceleration. If this is the relevant reference class for evaluating the plausibility of AI-driven explosive growth, we ought to assign a low prior chance to the possibility of explosive growth driven by AI.


Our view is that this argument is sound in general and gives us some uninformative prior over whether any new technology is likely to lead to explosive growth. The probability of this happening for a generic technology is, indeed, quite small: for instance, while fusion reactors would no doubt be economically valuable, we do not expect them to lead to explosive growth even if they became viable and cost-effective. However, the evidence that AI that can match human performance on most or all economic tasks is likely to lead to explosive growth is strong enough to overcome this general argument.

The key reason is that almost every model in endogenous growth theory makes the prediction that AI that is capable of automating most or all economic tasks humans can perform at low cost (e.g. cost of human subsistence) has a substantial chance of leading to explosive growth. For some models, this prediction is robust to parameter choices; while in others it is sensitive, but in either case we cannot rule out the possibility. For example, \hyperref[sec:increasing-returns-to-scale]{Section 2.1} predict explosive growth robustly conditional on full automation from AI, while, as we show in \hyperref[sec:argument-computing-costs]{Section 2.2}, constant returns to scale models make this prediction for a substantial fraction of plausible parameter values.

There is no comparable situation with most other technologies, and the reason is the important role played by labor in growth economics. Labor is unique in that it is an input that is \textit{both} a key driver of economic production and growth \textit{and} cannot be increased by reinvestment of economic output the same way capital, compute, energy etc. production can be. In other words, labor is \textit{non-accumulable}, while other factors of production that are of comparable importance to labor are \textit{accumulable}.

This means the potential economic benefits of a technology that can turn labor into an accumulable input are enormous: we turn the currently most important factor of production from something that is difficult to scale to something that is easy to scale. If we also assume that the cost of producing or maintaining this stock of accumulable labor inputs is not prohibitively expensive, almost all conventional growth models will predict explosive growth in this situation.


While the generic argument outlined in this section is convincing about most technologies, we believe that in the specific case of AI that is capable of substituting for human workers, we have enough evidence to overcome the low prior that such an argument would assign to explosive growth conditional on AI. As a result, if the other objections to our argument (regulations, other bottlenecks, slow speed of automation, etc.) do not apply, we think this generic argument does not have any additional force. For this reason, we think it is \textbf{very unlikely} that this argument blocks explosive growth.

\subsection{Fundamental physical limits restrict economic growth}

There might be fundamental physical limits to how much we can produce with a given amount of resources, or how quickly we can scale up production from current levels, regardless of how good our technology is. This objection may, for example, be found in (\cite{aghion2018artificial}). If these limits are sufficiently tight, they might prevent explosive growth. Some examples of such limits include the speed of light, conservation of energy, the Landauer limit for irreversible computing, the Bekenstein bound for energy density, Bremermann's limit for reversible computing, Carnot’s theorem, etc.


In principle, this argument is valid: there will be fundamental physical limits that block economic growth at some point. Many, if not all, of the bounds listed above will be relevant in constraining growth in the far future. However, we find the argument unconvincing insofar as it is meant to apply to explosive growth caused by AI automation this century, because we're simply too far from the relevant fundamental physical limits for the constraints imposed by them to be binding.

For instance, humans use around \( 0.01 \% \) of the energy flux incident on Earth for production and consumption, and doing \( 10^{40} \flopyr \) of computation on Earth alone seems feasible based only on fundamental physical limits, which at the cost of \( \sim 10^{23} \flopyr/\textrm{person} \) estimated in \cite{carlsmith2020} for running the human brain would be sufficient to simulate \( 10^{17} \) virtual workers. This would be equivalent to scaling up the world population by 7 orders of magnitude. Even if every worker needs to be provided with amenities that match the current per capita energy consumption on the planet, there is still room for a scaling up of 3 to 4 orders of magnitude.

We simply cannot come up with any plausible scenario in which economic growth is blocked early on as a result of a fundamental physical limit, as opposed to e.g. limitations of our engineering capabilities. As a result, we think this argument is rather weak. We think the chance that this argument blocks explosive growth conditional on widespread AI automation is small and conclude that it is \textbf{very unlikely} to block explosive growth.

\section{Discussion}

Our analysis of the potential for explosive growth from widespread AI automation reveals several key insights. The most significant finding is that economic growth models consistently predict explosive growth when AI can effectively replace human labor across most or all economic tasks. This prediction is robust across various types of models, including semi-endogenous and exogenous growth models, those with increasing, constant or even decreasing returns to scale, with ideas becoming `harder to find' and in models that account for investment delays. The consistency of this prediction across different model specifications strengthens the case for considering AI-driven explosive growth as a serious possibility.

By contrast, we find that many of the arguments against explosive growth lack quantitative specificity or are found to be relatively weaker. For instance, if we grant that regulation or human-preferences for human-produced goods create large pockets of economic activity that will be non-automated, then a `Baumol' style bottleneck arguments against explosive growth is still relatively weak in light of the massive transitory gains in output that these still permit. Moreover, it is unclear whether these bottlenecks will last for prolonged periods of time, as this might require prolonged and widespread regulation against the backdrop of potentially enormous gains from expanding AI's deployment.

Having gone through the above arguments for and against explosive growth, we think that explosive growth by the end of this century is about as likely as not conditional near-complete AI automation.\footnote{While we don't make the case for the unconditional view here, we separately think this is plausible based on estimates of how much resources would be needed for the creation of an AI capable of matching or surpassing human performance across most tasks and how much we can expect effective investment into AI to get scaled up by the end of this century. This case is made in greater detail elsewhere, e.g. \cite{cotra2020forecasting}, \cite{davidson2023compute} and \cite{epoch2023thedirectapproach}. We do not reproduce the detailed arguments here and direct the interested reader to these more comprehensive sources.} 

Due to the numerous arguments against this conclusion that we have discussed here and the prediction of explosive growth involving the extrapolation of models beyond the regime in which they have been observed to work, we think high confidence in explosive growth is unwarranted. However, we think confidence in explosive growth not occurring is also unwarranted, especially conditional on the arrival of AI capable of substituting for human workers. 

The distinct arguments for and against explosive growth are likely correlated with each other so their disjunction is less likely than we might otherwise infer under an independence assumption. This has important implications for our assessment. There may be unknown variables (How economically valuable is a large increase in AI workers? What is the economic value of superhuman intelligence?) that simultaneously influence multiple arguments. If these latent factors tend in a certain direction, they could make several arguments for (or against) explosive growth more likely to be true concurrently. In other words, the conjunction of many arguments are more likely than than we might otherwise infer under an independence assumption.\footnote{To formally illustrate the point about correlations, we can compute:
\(\mathbb P \left( \bigcup_{i=1}^n A_i \right) = 1 - \mathbb P \left( \bigcap_{i=1}^n A_i^c \right) = 1 - \mathbb P(A_1^c) \prod_{i=2}^n \mathbb P(A_i^c \vert A_1^c, \ldots, A_{i-1}^c),
\)
where the superscript \( c \) denotes taking complements and \( A_1, \ldots, A_n \) are \( n \) events on a probability space. When the arguments are correlated with each other, \( \mathbb P(A_i^c \vert A_1^c, \ldots, A_{i-1}^c) > \mathbb P(A_i^c) \), so the product is larger than it would be were the events jointly independent, and the probability of the disjunction is accordingly smaller.}

For example, if AI that substitutes for human workers has smaller economic impacts than economic models predict, we might see a cascade of related effects. Automation could be more drawn out due to reduced economic incentives for investment and weaker feedback effects from AI automation. Regulating AI might face fewer obstacles because of the reduced economic windfall from AI deployment. AI alignment issues might be resolved more slowly due to less perceived urgency. In other words, a single underlying factor can influence multiple aspects of AI development and deployment, potentially reinforcing a particular outcome.

We think the most plausible such underlying factor that would induce such a correlation is the ``overall level of problem-solving capabilities" of many human-level or superhuman intelligences operating in parallel for long durations of subjective time. The more powerful intelligence turns out to be in general, the more easily we will be able to find ways to get around bottlenecks, and so conditional on one bottleneck not being serious, the likelihood of others being serious goes down as well.

There are other confounding influences as well. For instance, many of the arguments that function through the channel of ruling out the economic equivalent of full automation rely on the elasticity of substitution parameter \( \sigma \) being small, meaning that it is harder to substitute goods from non-automated sectors with goods by AI. That said, such arguments might require this parameter to be smaller than values often reported in the literature for the elasticity of substitution between capital and labor e.g. in the US economy, for instance in \cite{knoblach2020elasticity}. Yet, this influence similarly lowers the probability of the disjunction of all arguments that depend on this key parameter.

After taking both the individual strength of the arguments and their overall correlation structure into account, we end up thinking that credences of less than \( 20 \% \) for explosive growth conditional on AI that can do most or all tasks in the economy are unreasonably low. We estimate \( \mathbb P(\text{explosive growth this century} \mid \text{AGI this century}) \) \textbf{about as likely as not}.

\subsection{Open questions}

There are several important questions that would make us more or less confident in explosive growth. Below is a non-exhaustive list of these questions:

\begin{enumerate}
    \item Are there competing theories of economic history that are similarly plausible to the semi-endogenous growth story? What do these alternative theories have to say about the deployment of AGI?
    
    \item How does the value-add of a technology affect the strength of regulation of coordination preventing its deployment? Do regulatory-induced delays follow a power law with respect to the value of relevant technologies? Is there strong evidence that innovations whose value is on the order of a ten times increase in the GDP of frontier economies are often blocked for a long time, i.e. many decades?

    \item How expensive will it be to build robotic systems for AGIs with adequate motor control to do most or all embodied economic tasks humans are able to perform? Will robotics costs be of the same order of magnitude as compute costs, lower or higher? Note that economies of scale are likely to be quite important here, so looking at present robotics costs could be misleading.

    \item Are early AI alignment failures going to make the deployment of otherwise capable AI systems by private actors unprofitable? While it is often assumed that misaligned AI would be deceptive and do what you want early on before it is sufficiently capable, leading to a situation in which actors who care about safety have to pay an ``alignment tax", in our view this position is not supported by strong enough evidence for us to simply take it for granted.

    If AI becomes so unsafe that deployment is in expected value sufficiently costly even from a private actor's point of view, then slowing AI down becomes a matter of self-interest and not of global coordination, which is important for assessing the likelihood of large slowdowns actually occurring.

    \item In our analysis, we consider both land and energy as physical factors which could bottleneck production. In both cases, we find that fundamental physical limits are at least some orders of magnitude away from our current use of these factors. Is this analysis flawed? If not, are there other factors that we have neglected which could similarly bottleneck output and prevent explosive growth?

    \item What is the economic value of superhuman intelligence? To make this question quantitative in one way (though certainly not the only way), how much more economically valuable would a human be if they had a brain that was ten times larger or faster, and how much more overhead in energy and other costs would this incur in humans? How favorable is this scaling relationship once we take both economic benefits and costs into account?

    For instance, a rough intuition here could be that a brain twice as large is roughly four times as economically valuable, though this kind of scaling could be quite naive for many reasons.
\end{enumerate}

All of these questions, if answered, could affect our views considerably. For instance, if superhuman intelligence is extremely powerful, then our credence in explosive growth this century should go up, as substantial expansions in our compute stock may not be needed for explosive growth. Some of these questions seem quite difficult to answer, while other questions seem amenable to progress. For instance, an in-depth investigation into the economics of robotics could plausibly answer (3), and a review of economic history from a quantitative lens could shed some light on (2). 

\printbibliography

\section*{Appendix}

\subsection*{Appendix A: Likelihood scale}
\label{sec:appendix-a}

To communicate our uncertainty appropriately, use the following likelihood scale in our assessment of the likelihood of explosive growth from AI occurring or being undercut by any of the obstacles discussed. 

\begin{table}[h]
\centering
\begin{tabular}{@{}p{5cm}p{5cm}@{}}
\toprule
\centering Term & \centering Likelihood of outcome \tabularnewline
\midrule
\centering Virtually certain & \centering \textgreater{}99\% probability \tabularnewline
\centering Very likely & \centering 90\%-99\% probability \tabularnewline
\centering Likely & \centering 66\%-90\% probability \tabularnewline
\centering About as likely as not & \centering 33\%-66\% probability \tabularnewline
\centering Unlikely & \centering 10\%-33\% probability \tabularnewline
\centering Very unlikely & \centering 1\%-10\% probability \tabularnewline
\centering Exceptionally unlikely & \centering 0\%-1\% probability \tabularnewline
\bottomrule
\end{tabular}
\caption{\small \centering Likelihood scale.}
\end{table}

\subsection*{Appendix B: Semi-endogenous growth models and idea production}
\label{sec:appendix-semi-endog}

This appendix contains some technical details on the high-level argument laid out in the \hyperref[sec:increasing-returns-to-scale]{increasing returns to scale section}.

First, let's at a high level why we should expect hyperbolic growth to occur when accumulable inputs have increasing returns to scale in the production function.

Suppose that \( Y: (\mathbb R^{\geq 0})^n \to \mathbb R^{\geq 0} \) is a production function mapping factor inputs \( (f_1, f_2, \ldots, f_n) \) (which might be labor, capital, etc.) to economic output \( Y \). If all of these inputs are strictly accumulable, in the sense that they can be increased proportionally by reinvestment of output \( Y \), then if we assume a fraction of output \( \alpha_k \) is invested in the accumulation of input \( f_k \), these quantities will satisfy the differential equations
\[ \frac{d f_k}{dt} = \alpha_k Y \]
For technical reasons that will become apparent soon, we want to choose the saving rates \( \alpha_k \) such that the factor ratios \( f_i/f_j \) are held constant. This is equivalent to choosing \( \alpha_i \propto f_i \), so if the overall saving rate of our economy is \( 0 < \alpha < 1 \), we'll have
\[ \alpha_i = \frac{\alpha \cdot f_i}{\sum_j f_j} \]
Using the chain rule on the production function \( Y \) gives
\[ \frac{dY}{dt} = \sum_{k=1}^n \frac{\partial Y}{\partial f_k} \frac{df_k}{dt} = \sum_{k=1}^n \frac{\partial Y}{\partial f_k} \alpha_k Y \]
and a substitution of the above expression for \( \alpha_i \) to this expression yields
\[ \frac{dY}{dt} = \alpha \times \sum_{k=1}^n \frac{\partial Y}{\partial f_k} \frac{f_k Y}{\sum_j f_j} \]
Now, we bring in the assumption that \( Y \) has increasing returns to scale. Suppose that \( Y \) is homogeneous of degree \( d > 1 \), so that it satisfies the homogeneity identity
\[ Y(rf_1, rf_2, \ldots, rf_n) = r^d Y(f_1, f_2, \ldots, f_n) \]
for all nonnegative real numbers \( r \). Since we assume factor ratios are held constant, our factor vector will always be of the form \( (rh_1, rh_2, \ldots, rh_n) \) for some real number \( r \) and our initial factor endowments \( h_1, h_2, \ldots, h_n \). Since \( Y \propto r^d \) by the above identity and \( \sum_i f_i \propto r \) by assumption, in particular we deduce that \( \sum_i f_i \propto Y^{1/d} \). Substituting into the above relation for \( dY/dt \) gives
\[ \frac{dY}{dt} \propto Y^{1 - 1/d} \sum_{k=1}^n \frac{\partial Y}{\partial f_k} f_k \]
Finally, suppose we differentiate the homogeneity identity for \( Y \) with respect to \( r \) at \( r = 1 \), holding factor inputs fixed. This gives the relation
\[ \sum_{k=1}^n f_k \frac{\partial Y}{\partial f_k} = d \times Y \]
Using this relation as the final ingredient, we get that \( Y \) satisfies a differential equation
\[ \frac{dY}{dt} \propto Y^{2 - 1/d} \]
exactly as claimed in the \hyperref[sec:increasing-returns-to-scale]{increasing returns to scale section}. When \( Y \) has increasing returns to scale, so that the homogeneity degree \( d > 1 \), \( Y \) exhibits hyperbolic growth and diverges in finite time. We also see why a transition in which a particular input shifts from being accumulable to not being accumulable can lower \( d \) and as a result shift us from a superexponential to a subexponential growth regime.

One important detail here is that the saving rule we chose for our economy, that \( \alpha_k \propto f_k \), is not necessarily optimal. However, the fact that \textit{some} saving rule can achieve hyperbolic growth is a sufficient condition for the economy to exhibit hyperbolic growth in the absence of severe market failures, so this is not an important issue.

\subsubsection*{Diminishing returns in factor production}
\label{sec:appendix-diminishing-returns}

This simple story is complicated when we consider more general laws of motion for the factors of production \( f_i \). \cite{bloom2020} considers a general accumulation relationship
\[ \frac{1}{f} \frac{df}{dt} \propto f^{-\phi} I^{\lambda} \]
where \( f \) denotes factor stock and \( I \) denotes investment into increasing this factor. In this formalism, the quantity \( r = \lambda/\phi \) (sometimes called the \textit{returns on factor investment}) is of crucial importance, as it determines the relationship between the growth rate of \( I \) and the growth rate of \( f \) in an exponential growth equilibrium.

It turns out we can generalize the above argument to the case where each factor follows an individual law of motion
\[ \frac{1}{f_i} \frac{df_i}{dt} \propto f_i^{-\phi_i} I_i^{\lambda_i} \]
but our result ends up being not quite as sharp. If we assume as before that factor ratios must stay constant, it follows that we must have
\[ f_i^{-\phi_i} I_i^{\lambda_i} \propto f_j^{-\phi_j} I_j^{\lambda_j} \]
for all \( i, j \). It straightforwardly follows that we must have \( I_i \propto f_i^{\phi_i/\lambda_i} = f_i^{1/r_i} \) where \( r_i = \lambda_i/\phi_i \) is defined as above, and the budget constraint \( \sum_i I_i = \alpha Y \) once again gives
\[ I_i = \alpha Y \times \frac{f_i^{1/r_i}}{\sum_i f_i^{1/r_i}} \]
As before, differentiating \( Y \) and using the chain rule gives us
\[ \frac{dY}{dt} = \alpha \times \sum_{k=1}^n f_k \frac{\partial Y}{\partial f_k} \frac{Y}{\sum_j f_j^{1/r_j}} \]
The problem is that when the \( r_j \) are different and the ratios between the different \( f_j \) are fixed by assumption, the denominator here will be dominated by the factor with the least favorable returns to investment. In other words, the best we can do is to bound the denominator from above using the relation
\[ \sum_j f_j^{1/r_j} = O(Y^{1/(d \min \{r_1, r_2, \ldots, r_n \})}) \]
Denoting \( r_{\text{min}} = \min \{r_1, r_2, \ldots, r_n \} \), we can obtain a lower bound on the growth of \( Y \):
\[ \frac{dY}{dt} >_{\text{up to a constant}} Y^{2 - 1/(d r_{\text{min}})} \]
As before, this is merely a sufficient condition, not a necessary one. However, if we make no further structural assumptions about \( Y \), this bound is the best we can do: assuming that \( Y \) is a Leontief production function, for instance, gives a concrete case in which we must keep factor endowments proportional to each other, so this worst-case bound ends up being tight. To relax this worst-case bound, it is necessary for the factors to not be perfect complements to each other.

It is also necessary to relax this bound if we hope to get explosive growth out of the argument. In \cite{bloom2020}'s formalism, the returns to idea production are by assumption equal to \( 1 \) (without loss of generality), so if accumulable inputs also have constant returns to scale we will have \( d = 2 \). In such a situation, we'll get explosive growth unconditionally if \( r_{\text{ideas}} > 1/2 \). However, \cite{bloom2020} estimates \(  r_{\text{ideas}} \approx 0.32 \) for the whole US economy, so this weak sufficient condition alone is insufficient to deduce we will have explosive growth once labor becomes accumulable.

\subsubsection*{Focusing on idea production}

Fortunately for us, the above calculation \textit{is} in fact too general, at least from the point of view of \cite{bloom2020}. This is because in their model, ideas enter the production function as a constant multiplier, meaning that we can narrow down the production function of the economy to a more specific form
\[ Y(A, f_1, \ldots, f_n) = A Y_f(f_1, f_2, \ldots, f_n) \]
where \( A \) represents total factor productivity and \( f_1, \ldots, f_n \) are accumulable factors as before. \( Y_f \) is also assumed to be homogeneous of degree \( d \). We have the laws of motion
\[ \frac{df_i}{dt} = I_i \]
\[ \frac{1}{A}\frac{dA}{dt} \propto A^{-\phi} I_A^{\lambda} \]
We now assume that we follow the previous investment allocation rule for \( Y_f \), so the ratios between the accumulable factors \( f_i \) are preserved, but unlike in the \hyperref[sec:sec:appendix-diminishing-returns]{diminishing returns in factor production section} we exclude \( A \) from the set of factors among which ratios must be preserved. Instead, we assume that a share \( \alpha_A \) of GDP is invested into idea research, and a share \( \alpha_f \) is invested in aggregate into accumulable factors. Treating these as constants, this yields
\[ \frac{1}{Y} \frac{dY}{dt} = \frac{1}{A} \frac{dA}{dt} + \frac{1}{Y_f} \frac{dY_f}{dt} >_{\text{up to a constant}} A^{-\phi} Y^{\lambda} + Y Y_f^{-1/d} = A^{-\phi} Y^{\lambda} + A^{1/d} Y^{1-1/d} \]
Now, let \( x = 1/(1 + d \phi) \). Note that \( 0 < x \leq 1 \). Our idea is to simplify the expression using the weighted arithmetic-geometric mean inequality 
\[ xa + (1-x) b \geq a^x b^{1-x} \]
using \( x \) as our relative weight between the two terms, which holds whenever \( a, b \) are both positive and \( 0 \leq x \leq 1 \). So we write
\[ \frac{1}{Y} \frac{dY}{dt} >_{\text{up to a constant}} A^{-\phi} Y^{\lambda} + A^{1/d} Y^{1-1/d} > x A^{-\phi} Y^{\lambda} + (1-x) A^{1/d} Y^{1-1/d} \]
and use the weighted arithmetic-geometric mean inequality mentioned above to obtain
\[ \frac{1}{Y} \frac{dY}{dt} >_{\text{up to a constant}} A^{-\phi x + (1-x)/d} Y^{\lambda x + (1-1/d)(1-x)} \]
By our choice of the value of \( x \), the exponent of \( A \) is equal to zero, so it drops out of the expression altogether. Substituting \( x = 1/(1 + d \phi) \) in the exponent of \( Y \), we can simplify the right hand side to obtain
\[ \frac{1}{Y} \frac{dY}{dt} >_{\text{up to a constant}} Y^{(r+d-1)/(\phi^{-1} + d)} \]
As the denominator of the exponent is always positive, it follows that \( Y \) exhibits hyperbolic growth and diverges in finite time whenever \( r+d > 1 \). It is easy to see that there is also an equilibrium where \( Y \) grows exponentially when \( r+d = 1 \), so this condition is both necessary and sufficient for explosive growth in this model.

As we mentioned earlier, the data from \cite{bloom2020} suggests \( r \approx 0.32 \), which means that the returns to scale on accumulable inputs can be as small as \( d \approx 0.68 \) while still leaving open the possibility of explosive growth.

\subsection*{Appendix C: Bounds on human population growth explain limits of historical growth in semi-endogenous growth models}
\label{sec:appendix-bounds}

Human population growth is bounded from above by biological constraints on human reproduction. That is, $L$ cannot grow faster than some rate $\bar{n}$. If so, semi-endogenous growth theory predicts a bound on economic growth that of a similar order as $n$. To see this, consider a semi-endogenous growth model described by the following equations:
\begin{align}
    Y(t) &= A(t)\big(K(t)\big)^{\alpha} \big((1-\alpha_l)L(t)\big)^{1-\alpha} \\
    \dot{A}(t) &= \alpha_l L(t)^\gamma A(t)^\phi \\
    \dot{K}(t) &= sY(t) -\delta K(t) \\
    \dot{L}(t) &= nL(t).
\end{align}
That is, we consider a simple semi-endogenous growth model with Hicks-neutral technical change, constant savings rate, and with scientists split between final goods production and R\&D. Solving the steady-state growth rates, we get that:
$$g_a = \frac{\gamma}{1-\phi}n, \, \, \, g_k = n \bigg[\frac{\gamma + (1-\phi)(1-\alpha)}{(1-\phi)(1-\alpha)}\bigg].$$
The steady-state rate of growth is thus:
$$g_y = n\bigg(\alpha\frac{\gamma + (1-\phi)(1-\alpha)}{(1-\phi)(1-\alpha)} + \frac{\gamma}{1-\phi} (1-\alpha)\bigg).$$
Hence, $g_y$ is proportional to $n$. For instance, if we follow the meta-analyses from \cite{sequeira2020stepping} and \cite{neves2018spillovers} and adopt $\phi = 0.8$, and $\gamma=0.2$, and as is standard, assume $\alpha = 0.3$, then $g_y \approx 1.5n$. Hence, semi-endogenous growth theory predicts that growth is capped at some rate that is, in some sense, quite close to $\bar{n}$.

\subsection*{Appendix D: Explosive growth from growth in stock of digital workers}
\label{sec:appendix-digital}

Consider an exogenous growth model with technological progress, where investment is split between compute and other capital:
\begin{equation*}
    Y(t) = A L(t)^{\alpha}  K(t)^{1-\alpha}, 
\end{equation*}
the stocks of effective labor and capital grow as a result of investment:
\begin{equation}
    \frac{dL(t)}{dt} = sfY(t)/\bar{c} - \delta_L L, \,
    \frac{dK(t)}{dt} = s(1-f)Y(t) - \delta_K K 
\end{equation}
where $f$ is the fraction of investment channelled towards AI, $ s $ is the saving rate of the economy, and $\bar{c}$ the average cost of running a human-equivalent AI. \( \delta_L, \delta_K \) are the depreciation rates for the effective labor and capital stocks, respectively. Assuming that $A$ is constant, some algebra reveals that:\footnote{The depreciation rate sets the timescale over which the hardware is useful: it is \( \sim 1/\delta_L \).}
\begin{equation}
    g_y = A s \left[ \alpha \left(\frac{K(t)}{L(t)}\right)^{1-\alpha} \frac{f}{\bar{c}} + (1-\alpha) \left(\frac{K(t)}{L(t)}\right)^{-\alpha} (1-f) \right] - \alpha \delta_L - (1-\alpha) \delta_K
\end{equation}
Along a balanced growth path, the ratio \( L/K \) should be equal to \( f/(1-f) \cdot 1/\bar{c} \). Substituting this into the expression for \( g_y \) and optimizing over \( f \) to find the value that leads to the highest growth rate in the long run gives \( f = \alpha \), so we can assume that after labor becomes accumulable, in the long run we will have \( L/K \approx \alpha/(1 - \alpha) \cdot 1/\bar{c} \).

Substituting into (2) would then lead to
\begin{equation}
    g_y = A s \frac{1}{\bar{c}^{\alpha}} B_\alpha
- \alpha \delta_L - (1-\alpha) \delta_K
, \,
B_\alpha = \left[ 
    \alpha^2 \left(\frac{1-\alpha}{\alpha}\right)^{1-\alpha}
    +
    (1-\alpha)^2 \left(\frac{\alpha}{1-\alpha}\right)^{\alpha}
\right] 
\end{equation}
We can also get an estimate for the value of \( A \) for a frontier economy. The total capital stock of the US economy \href{https://fred.stlouisfed.org/graph/?g=13rR9}{was estimated} in 2019 to be around 70 trillion USD in 2017 prices. Furthermore, the number of employed people in the US in 2019 was around 180 million: there were 150 million nonfarm employees \href{https://fred.stlouisfed.org/graph/?g=13maw}{according to FRED}, and the same page states that nonfarm employment makes up for around \( 80 \% \) of the employees that contribute to gross domestic product. Finally, \href{https://fred.stlouisfed.org/graph/?g=13rRx}{US real GDP} was around 19 trillion 2012 USD, or 20 trillion 2017 USD, in the year 2019.

Combining all of this information and assuming \( \alpha = 0.7 \) gives us the equation
\[ 2 \times 10^{13} \, \$/\text{year} = A \times (1.8 \times 10^8 \, \text{workers})^{0.7} \times (7 \times 10^{13} \, \$)^{0.3} \]
Solving for \( A \) yields
\[ A \approx 2337 \, \$^{0.7} \text{workers}^{-0.7} \text{year}^{-1} \]
We can now put all of this together to compute the growth rate we should expect post-AGI. If we make the simplification that \( \delta_L, \delta_K \ll g_y \), we can approximate the solution by
\begin{align}
    g_y &\approx A(t) s \bar{c}^{-0.7} \big( 0.7 \cdot (0.3/0.7)^{0.3} \cdot 0.7 + 0.3 \cdot (0.3/0.7)^{-0.7} \cdot 0.3 \big) \\
        &\approx 2337 \times s \bar{c}^{-0.7} \times 0.54 \\
        &\approx 1262 \times s \bar{c}^{-0.7}
\end{align}
where the units of the final answer will be year inverse. Explosive growth requires \( g_y \geq 0.3 \), suggesting the bound
\[ s \bar{c}^{-0.7} \geq 0.3/1262 \approx 2.38 \times 10^{-4} \]
or, cast slightly differently,
\[ \bar{c} \leq s^{10/7} \cdot (1.5 \times 10^5) \, \$/\text{worker} \]

\subsection*{Appendix E: Robustness to investment delays}
\label{sec:appendix-robustness}

\small
Here, we show that our earlier results are robust even if we assume that there are delays to investment, in the sense that ``realized investment" is an exponential moving average of past inputs to investment. Formally, with a ``forgetting rate" of \( \eta > 0 \), such a model would look like
\begin{align}
    \frac{dK}{dt} &= I \\
    \frac{dI}{dt} &= \eta(sY - I) \\
    Y &= AK
\end{align}
Here, \( s \) is a constant factor saving parameter as above, \( A \) is a multiplier with dimensions of frequency that converts the capital stock (which has dimensions of dollars) into GDP (which has dimensions of dollars per unit time), and \( \eta \) is a parameter with dimensions of frequency that controls how responsive realized investment \( I \) is to changes in savings \( sY \).

As we shall see,  \( 1/\eta \) is the ``characteristic time scale" of investment delays in this model. Specifically, \( \log(2)/ \eta  \) is the number of months for investment to move halfway between I and sY.

This is a straightforward system of differential equations and the asymptotic growth rate will be determined by the positive eigenvalue of the associated matrix
\[ A = \begin{bmatrix}
    0 & 1 \\
    As \eta & -\eta
    \end{bmatrix}
\]
The characteristic polynomial of this matrix is \( \det(tI - A) = t^2 + \eta t - As \eta \), which has positive root
\[ \lambda = \frac{-\eta + \sqrt{\eta^2 + 4As \eta}}{2} = \eta \times \frac{-1 + \sqrt{1 + 4As/\eta}}{2} \]
When \( \eta = \infty \) so that adjustment is instant, the steady state growth rate should be \( As \). Indeed, this is true in the limit, as can be seen from the first order approximation \( \sqrt{1 + \varepsilon} = 1 + \varepsilon/2 + O(\varepsilon^2) \).

Quantitatively, the deviation from this limit is insignificant unless \( \eta \ll 4As \). The order of magnitude here is dominated by \( A \) as \( s \) is a dimensionless saving rate parameter and \( 4 \) is a constant, so roughly speaking this expression is comparing the two frequency parameters \( A \) and \( \eta \). In the calculation from the previous section, the constant that corresponds to \( A \) is \( 2/3 \, \text{years}^{-1} \), so this means that we won't see a substantial impact of delays to investment on the growth rate of the economy unless \( \eta \ll 2/3 \, \text{years}^{-1} \) or \( 1/\eta \gg 18 \, \textrm{months} \). 

The most likely values for \( 1/\eta \), which is the ``characteristic time scale" of investment delays in this model, are probably on the order of a few years. Therefore this simple calculation predicts a constant factor effect of investment delays on the growth rate of the economy after full automation. Indeed, if we assume \( \eta = As \), this constant factor is 

\[ \frac{\sqrt{5} - 1}{2} = \frac{1}{\phi} \approx 0.618 \ldots \]

which is the reciprocal of the golden ratio, meaning that the growth rate is reduced to roughly \( 60 \% \) of what we would have predicted it to be in the naive model not taking these adjustment costs into account. Overall, we think the uncertainty this adds to the calculation is much smaller than the uncertainty already present from our estimates of \( A \) and \( s \), so this effect looks like it can be safely ignored, perhaps at the expense of choosing other model parameters a bit more conservatively.

\end{document}